\documentclass[onecolumn,amssymb, nobibnotes, aps, prfluids]{revtex4-1}

\usepackage{graphics}
\usepackage{bm}
\usepackage{graphicx}
\usepackage{amsmath}
\usepackage{amssymb}
\usepackage{float}
\usepackage{mathtools}
\usepackage{subfigure}

\newcommand{\be}{\begin{equation}}
\newcommand{\ee}{\end{equation}}
\newcommand{\bea}{\begin{eqnarray}}
\newcommand{\eea}{\end{eqnarray}}

\begin{document}

\title{Fluctuations of Lyapunov exponents in homogeneous and isotropic turbulence}

\author{Richard D. J. G. Ho}
\email{richard.ho@ed.ac.uk}
\author{Andr\'es Arm\'ua}
\email{andres.armua@ed.ac.uk}
\author{Arjun Berera}
\email{ab@ph.ed.ac.uk}
\affiliation{
School of Physics and Astronomy,
University of Edinburgh, Edinburgh EH9 3FD, United Kingdom}

\begin{abstract}
In the context of the analysis of the chaotic properties of 
homogeneous and isotropic turbulence, direct numerical simulations are used to study the fluctuations of the finite time Lyapunov exponent (FTLE) and its relation to Reynolds number, lattice size and the choice of the steptime used to compute the Lyapunov exponents.
The results show that using the FTLE method produces Lyapunov exponents that are
remarkably stable under the variation of the steptime and lattice size. Furthermore, it reaches such stability faster than other characteristic quantities such as energy and dissipation rate.
These results remain even if the steptime is made arbitrarily small.
A discrepancy is also resolved between previous measurements of the dependence on the Reynolds
number of the Lyapunov exponent. The signal produced by different variables in the steady state is analyzed and the self decorrelation time is used to determine the run time needed in the simulations to obtain proper statistics for each variable. 
Finally, a brief analysis on MHD flows is also presented, which shows that the Lyapunov exponent is still a robust measure in the simulations, although the Lyapunov exponent scaling with Reynolds number is significantly different from that of magnetically neutral hydrodynamic fluids.  
\end{abstract}

\date{\today}

\maketitle

\section{Introduction}
\label{S:HD}

It is well-known that turbulence implies chaotic behaviour \cite{ruelle1971nature,deissler1986navier}. The description of chaos in turbulent flows has a wide range of applications ranging from geophysical to engineering flows \cite{lorenz1963deterministic,yoden1993finite,masuoka2003chaotic,musker1979explicit,poirel2001structurally}. In particular, the chaotic study of turbulence is strongly associated with the problem of predictability \cite{leith1971atmospheric,leith1972predictability,lorenz1963deterministic,aurell1997predictability,yoden1993finite}. Chaos theory has also been applied to plasmas and astrophysical flows \cite{kurths1986can,grappin1986computation,huang1994estimation}. In the past few decades, there has been an increasing interest in numerically determining the degree of chaos in turbulent flows. 

In recent times, high performance computing has made possible to study chaotic properties of 3D flows using direct numerical simulations (DNS) \cite{Berera2018, Boffetta2017, mohan2017scaling,ho2019chaotic,nastac2017lyapunov,mukherjee2016predictability,li2019superfast}. This consists in evolving the Navier-Stokes equations (NSE) using no modelling. Almost two decades ago, DNS in 2D flows were used to study predictability \cite{boffetta1997predictability,boffetta2001predictability}. Chaotic properties of 3D turbulence have been also studied using approximate models such as EDQNM closure approximation \cite{metais1986statistical} or shell models \cite{Crisanti1993,aurell1997predictability,yamada2007chaotic}. All these works have studied chaos and predictability using the Eulerian description of fluids. A different approach commonly found in literature uses the Lagrangian description of turbulent flows to describe chaos \cite{deDivitiisNicola2018SoFS,lapeyre2002characterization,biferale2005lagrangian,biferale2005multiparticle}.

Whereas the Lagrangian approach deals with dispersion of pairs of fluid particles, 
the Eulerian approach measures the evolution of two different flows with similar initial conditions. The study of Lagrangian chaos is widely applied in geophysical \cite{yoden1993finite,vannitsem2017predictability,guo2016finite,garaboa2017climatology,d2004mixing,haller2011lagrangian,haller2015lagrangian,abraham2002chaotic,garaboa2017climatology} and astrophysical applications \cite{chian2019supergranular,rempel2016objective, chian2014detection} as well as the study of engineering flows \cite{tang2012lagrangian,pedersen1996lyapunov} and plasmas \cite{padberg2007lagrangian,falessi2015lagrangian,misguich1987diffusion}. In these contexts, chaos is used to characterize mixing and transport properties as well as determining Lagrangian coherent structures (LCS) \cite{shadden2005definition}. Unlike Eulerian chaos, several statistical analyses of Lagrangian chaos have been done in the past \cite{lapeyre2002characterization,deDivitiisNicola2018SoFS,johnson2015large}.

Eulerian chaos measures the evolution in state space of two initially close realizations 
of the velocity field $\bm{u} \equiv \bm{u}\left(\bm{x},t\right)$ 
and $\bm{u}+\delta \bm{u}$, where $\delta\bm{u}$ is infinitesimal. Given the chaotic nature, the difference between both realizations grows exponentially. This can be expressed as
\begin{equation}
\lvert \delta \bm{u}\left(t\right) \rvert \sim \lvert \delta \bm{u}\left(0\right)\rvert \exp \left(\lambda t\right)   \quad .
\label{eq:e_vs_lyap_t}
\end{equation}
Here $\lambda$ is the maximal Lyapunov exponent. This exponent is a measure of the level of chaos in a turbulent flow \cite{shimada1979numerical}. In other words, it sets the timescale in which the evolution of the flow can be predicted.

Lagrangian chaos statistics have been widely studied compared to the Eulerian case. For the Eulerian case, the method used in \cite{Boffetta2017,mohan2017scaling} produces several chaotic realizations that produces different Lyapunov exponents as the flow evolves, this allows to perform a statistical study of Lyapunov exponents. In \cite{mohan2017scaling}, it is noted that the time evolution of such Lyapunov exponent suffers large and fast fluctuations which differ from the fluctuations usually observed in other quantities such as energy or Reynolds number. Hence, a deeper study of the fluctuations of the Lyapunov exponents and their statistical properties is needed. 
Within a turbulent system there can be two types of fluctuations. One is of stochastic nature, it occurs due to the interaction with noisy environments such as molecular collisions in the small scale. The second type is given by the chaotic nature of the evolution in the NSE due to the non-linear term. These are called deterministic fluctuations. In this work we focus on describing deterministic fluctuations. Unlike experiments, numerical simulation do not introduce unwanted stochastic effects unless deliberately implemented. Thus, in this work, every fluctuation is deterministic.

There are a couple of reasons why we study the fluctuations in the Lyapunov exponents in the Eulerian description of a turbulent flow. One is the magnitude and timescales of such fluctuations are significantly different from other characteristic quantities such as energy, Reynolds number and dissipation. This shows the need to understand the properties of the flow that are particular to chaos and their relation to other parameters in the flow. The other is characterizing the distributions of flow quantities such as Reynolds number, dissipation rate or Lyapunov exponent is the key in any statistical description of turbulence. Applications may be possible since the stability of any quantity may be useful to characterize or calibrate systems. 

In 1979, Ruelle predicted that this exponent should be proportional to the inverse of the smallest timescale in the system, that is the Kolmogorov time $\tau = \left(\nu/\varepsilon\right)^{1/2}$, 
where $\varepsilon$ is the dissipation rate and $\nu$ 
the kinematic viscosity \cite{ruelle1979microscopic}.
Ruelle's prediction uses simple arguments based on dimensional analysis.
First, Ruelle states that the maximal Lyapunov exponent should be associated with the smallest 
timescale of the system, $\tau$.
Subsequent work related $\lambda$ to the Reynolds number, $\text{Re} = UL/\nu$, where $U$ is the rms velocity and $L$ the integral length scale. 
Here $L = (3\pi/2E)\int E(k) / k \ dk$, $E$ being the turbulent kinetic energy
and $E(k)$ the energy spectrum.
Using Kolmogorov (K41) theory \cite{KolmogorovA.N.1991DoEi}, 
it can be deduced that $\lambda \propto (1/T_0) \text{Re}^{1/2}$, 
where $T_0 = L/U$ is the large eddy turnover time. 

The relation $\lambda \sim 1/\tau$ proposed by Ruelle is based on very simple considerations. 
Nevertheless, in practice it is observed that the product $\lambda \tau$ is not constant, but it tends to increase 
slightly with Reynolds number \cite{Boffetta2017,Berera2018}. 
Using a multifractal model, and considering intermittency, the following scaling relation is derived \cite{Crisanti1993}
\begin{equation}
\lambda = \frac{D}{T_0} Re^{\alpha} \quad ,
\label{eq:ly_vs_Re^alph}
\end{equation}  
where $\alpha = (1-h)/(1+h)$, and $h$ is the H\"older exponent obtained from the 
structure function $\langle \lvert \bm{u}\left(\bm{x} +\bm{r}\right) - 
\bm{u}\left(\bm{x}\right)\rvert\rangle \sim V l^{h}$, where $l = r/L$. 
In Kolmogorov theory, $h=1/3$ so $\alpha = 1/2$ and Ruelle's scaling is recovered. 
Using a multifractal model of turbulence \cite{Crisanti1993}, 
the theory predicts a value of $\alpha < 0.5$, 
whereas numerical results show values of $\alpha > 0.5$ \cite{Boffetta2017,Berera2018}.

The paper is organized as follows. In section \ref{se:Method}, we describe the numerical method used in the simulations as well as the method used to compute the finite time Lyapunov exponents (FTLEs) and flow parameters. In section \ref{se:Hydro_Results}, we analyze the distributions obtained for Reynolds number and FTLEs (\ref{se:distributions}). Then we look at the relation of the Lyapunov exponent and its fluctuations to lattice size (\ref{se:lattice}), Reynolds number (\ref{se:Re_dependence}), and the steptime in the FTLE method (\ref{se:Steptime_dependence}). This analysis is done for a hydrodynamic flow evolved using the incompressible NSE . A discrepancy is noted between previous measurements by different authors \cite{Boffetta2017,Berera2018}. This is resolved by considering corrections given by the dimensionless dissipation rate \cite{Mccomb2015} (\ref{se:def_of_T_0}). Also for the case of NSE, we perform further analysis for chaos in decaying turbulence. This helps to determine which is the proper timescale to use in the study of Eulerian chaos (\ref{se:decaying}). 

 In section \ref{se:timescales_HIT}, we analyze the timescales in HIT. This is motivated by the fact that Lyapunov exponent signal presents a faster timescale than those of energy and dissipation rate. We make comparisons between different quantities such as energy, dissipation rate and Lyapunov exponents by looking at the decorrelation time in their signals. As a result, we give a useful rule to determine the run time needed to obtain useful statistics from knowledge of the decorrelation time in their signals. It is important to note that even though the analysis is motivated by the study of chaotic properties, it is not restricted to that case only but it is useful in any study of turbulence.

Our main finding is that the Eulerian maximum Lyapunov exponent is a robust measure
of the level of chaos in a turbulent flow
and gives stable statistics faster than other measurable quantities such as the Reynolds number or energy. Noting that relations like Ruelle's relate the Lyapunov exponent to other flow parameters such as the Reynolds number, we expect that this finding can be useful to characterize or test systems using the Lyapunov exponents \cite{nastac2017lyapunov}. Furthermore, we find that the measurements of FTLEs are stable even for very short steptimes, thus reducing the computational cost needed to obtain large sets and improve the statistics. 

In section \ref{se:MHD}, a brief analysis is done for Eulerian chaos in an incompressible MHD flow. We observe the distributions of Reynolds number and Lyapunov exponent, and we test the Ruelle's relation as well as the dependence of the FTLEs on the steptime in the FTLE method. We find that even though Ruelle's relation does not hold quantitatively for MHD, the Lyapunov exponent is still a robust measure that gives stable statistics faster than other quantities. Finally, in section \ref{se:discussions}, we discuss the ideas developed across the different sections and we present conclusions.  

\section{Method}
\label{se:Method}
In this work, several simulations evolving homogeneous and isotropic flows (HIT) were run using a 
pseudospectral method and a fully-dealiased DNS code in a periodic box 
with $N^3 = 32^3, 64^3, 128^3, 256^3$ and $512^3$ collocation points. Each flow is evolved using the incompressible NSE,
\begin{gather}
\partial_t \bm{u} = - \left(\bm{u} \cdot \nabla \right) \bm{u} + \nu \nabla^2 \bm{u}  + \bm{f} \ , \\
\nabla \cdot \bm{u} = 0 \quad ,  
\label{eq:incompressible}
\end{gather}
where $\bm{u}\left(\bm{x},t\right)$ is the velocity field 
and $\bm{f}$ represents an external force per unit volume. 
A negative damping forcing scheme is implemented. 
It consists in using the lowest modes in the velocity 
field $\hat{\bm{u}}\left(\bm{k},t\right)$ 
(the Fourier transform of $\bm{u}\left(\bm{x},t\right)$) 
as a forcing function,
\begin{equation}
    \hat{\bm{f}}(\bm{k},t) = 
    \begin{cases}
    \frac{\varepsilon}{2E_f} \, \hat{\bm{u}}(\bm{k},t) \quad &\text{if}\quad 0<\lvert\bm{k}\rvert \leq k_f \, , \\
        0 \quad  &\text{otherwise}  \quad , 
    
    \end{cases}
\end{equation}
where $E_f$ is the energy of the velocity field contained in the forcing bandwidth.
This well tested forcing function \cite{machiels1997predictability,ishihara2003high} has the
advantage that the dissipation rate $\varepsilon$ 
is a parameter that can be set \textit{a priori} in each simulation. 
In this case, $\varepsilon$ is set to $0.1$ for all simulations 
and $k_f = 2.5$, so only the large scales are forced. 
The field is initialized with a Gaussian distribution for components of the velocity
field including a random seed. A full description of the code and the forcing can be found in \cite{yoffe2013investigation}.

Once the fluid evolution reaches a statistically steady state, the averages of $L,T_0,U,\varepsilon$ and the finite time Lyapunov exponent $\lambda$ are computed
over multiple large eddy turnovers. 
However, a certain level of fluctuation is always observed.
This is possibly associated with the forcing mechanism together 
with the irregular spatio-temporal behaviour given by 
the highly non-equilibrium nature of the turbulent regime. 
A complete understanding and characterizations of these fluctuations 
are not the aim of this work, but it would be an interesting step to take in the future.
Instead, the main goal of this study is to understand the relations between these variables and also with other simulation parameters such as the lattice size $N$.
The finite time Lyapunov exponents are computed for each simulation using the method described in the following section.

As well as looking at the evolution of the NSE, we perform
a similar analysis for the case of homogeneous and isotropic MHD flows.
This analysis consists in evolving the velocity field $\bm{u}$ and the magnetic field $\bm{b}$ 
using the following incompressible MHD equations
\begin{align}
\label{eq:MHDeqns1}
    \partial_t \bm{u} &= - \left(\bm{u} \cdot \nabla \right) \bm{u} + \nu \nabla^2 \bm{u} + \left(\nabla \times \bm{b} \right) \times \bm{b} + \bm{f}  , \\
    \partial_t \bm{b} &= \left( \bm{b}\cdot\nabla\right) \bm{u} - \left(\bm{u} \cdot \nabla\right) \bm{b} + \eta \nabla^2 \bm{b} \quad , 
    \label{eq:MHDeqns2}
\end{align}
together with the incompressibility condition given by Eq.(\ref{eq:incompressible}) and  
\begin{align}
    \nabla \cdot \bm{b} = 0 \quad , 
\end{align}
where $\eta$ is the magnetic diffusion coefficient.
For MHD, the Lyapunov exponents measure the divergence of a combination of both the 
velocity and magnetic fields as described in the following section.

\subsection{Finite Time Lyapunov Exponents method}
\label{se:FTLE_method}

Finite time Lyapunov exponents are widely studied in many applications within the Lagrangian description \cite{yoden1993finite,lapeyre2002characterization, deDivitiisNicola2018SoFS,shadden2005definition,brunton2010fast}. According to the theory, the Maximal Lyapunov exponent do not depend on the initial infinitesimal perturbation \cite{oseledets1968multiplicative}, but a slight dependence is usually present for finite perturbations in finite times \cite{goldhirsch1987stability}. Hence, the Maximal Lyapunov exponent is estimated by averaging over an ensemble of different realizations of finite perturbations in finite times. 

In this section, we present the method used in this work to compute the FTLE, which is also used in \cite{Boffetta2017}. In the Eulerian description, the procedure used to measure Lyapunov exponents consists in
evolving two fields that are initially similar. Given the original field $\bm{u}_1$, another field $\bm{u}_2 = \bm{u}_1 + \delta \bm{u}$ is 
created. Then, we evolve the system over a finite steptime $\Delta t$,
measure the field difference,
and then introduce a successive perturbation, starting the process again each time.
The field difference is defined as 
\begin{align}
    \delta_t &= \sqrt{2 E_{ud}(t)} \quad , \\
    E_{ud} &= \frac{1}{2}\int d \bm{k} \lvert \hat{\bm{u}}_1\left(\bm{k},t\right) -  \hat{\bm{u}}_2\left(\bm{k},t\right) \rvert^2 \, .
\end{align}
The perturbation introduced every steptime $\Delta t$ is given by 
\begin{equation}
    \bm{u}_2 (\bm{x},\Delta t) = \bm{u}_1 (\bm{x},\Delta t) + \frac{\delta_0}{\delta_{\Delta t}} \, \delta \bm{u} \left(\bm{x},\Delta t\right)   \, ,
    \label{eq:steptime}
\end{equation}
where $\delta_0$ is set to  $10^{-3}$ and $\Delta t = 0.1$ except in Section \ref{se:Steptime_dependence}. 

Each exponent $\Tilde{\lambda}$ is measured as
\begin{equation}
    \Tilde{\lambda} = \frac{1}{\Delta t} \ln \left(\frac{\delta_{\Delta t}}{\delta_0}\right)  \quad  ,
\end{equation}
and this procedure is repeated, obtaining a set of values for $\Tilde{\lambda}$.
The mean Lyapunov exponent and its variance were calculated in the usual way
\begin{align}
\lambda &= \langle \Tilde{\lambda} \rangle \, , \\
 \sigma_{\lambda}^2 &= \frac{1}{n} \sum \langle \Tilde{\lambda}^2 \rangle -  \langle \Tilde{\lambda} \rangle ^2  \, ,
\end{align}
where $n$ is the number of FTLEs measured.

In the case of MHD, we use the same procedure. The only difference is that $\delta_t$ is taken as a combination of the increase in both the velocity and magnetic field difference
\begin{equation}
    \delta_t = \sqrt{2 E_{ud}(t) + 2 E_{bd}(t)} \quad , 
\end{equation}
where
\begin{equation}
E_{bd} = \frac{1}{2}\int d \bm{k} \lvert \hat{\bm{b}}_1\left(\bm{k},t\right) -  \hat{\bm{b}}_2\left(\bm{k},t\right) \rvert^2 .
\end{equation}

The advantage of the FTLE method is that it produces a sample of Lyapunov exponents 
that allows us to perform a statistical analysis of the measurements. 
Other methods exist in literature, for instance, the \textit{direct method} consists 
in introducing a small perturbation once the system reaches a steady state and then 
letting the two fields evolve until they become uncorrelated \cite{Berera2018,ho2019chaotic}. 
This method reveals interesting features because it displays the long time evolution 
of the spectrum of the field difference $\lvert \delta \bm{u}\rvert $ 
(in which a transient and a saturation stage are found \cite{Berera2018}).
However, this procedure takes a relatively large computational 
time to get just one Lyapunov exponent. 
In contrast, the FTLE method is useful to obtain a large number of Lyapunov exponents 
which can be used to analyze the statistical distribution for a given steady state. 
This procedure shows that fluctuations in the Lyapunov exponents are quite significant 
and this fact was not previously noted using the direct method. 
Hence, it is interesting to analyze the source of such fluctuations 
and see if they can be explained by other fluctuating quantities in the system.

\section{Hydrodynamic Results}
\label{se:Hydro_Results}
The values of $\lambda$ measured using the FTLE method produces a time signal with large fluctuations. These fluctuations are briefly mentioned in \cite{Boffetta2017} and some aspects are studied in \cite{mohan2017scaling,nastac2017lyapunov}.
In our work, we further analyze the properties of these signals. We look at the mean and standard deviation of $\lambda$ and study their dependence with different flow parameters to see if these variations are of physical origin or if they are caused by the choice of parameters in the numerical simulations.

First, we look at the probability distributions of Re and the FTLEs.
Second, we look at how the fluctuations in $\lambda$ depend on the lattice size and Reynolds number.
Third, we find a resolution to the discrepancy in results between those
of \cite{Berera2018} which quoted $\alpha = 0.53 \pm 0.03$ and \cite{Boffetta2017}, 
which had a higher value of $\alpha = 0.64 \pm 0.05$.
Fourth, we observe how the Lyapunov exponent fluctuations depend on the choice of the 
steptime parameter $\Delta t$ of the FTLE method.
Fifth, we study the chaotic behaviour of decaying turbulence and use the high sensitivity of the system with respect to the Lyapunov exponent to test which is the proper large timescale to describe Eulerian chaos. 

\subsection{Distributions of $\lambda$ and Re}
\label{se:distributions}

When the fluid reaches a steady state, its statistical quantities such as Re and total energy, 
fluctuate around a mean value. 
We first measure the fluctuations in Re, 
and then observe if these are consistent with the fluctuations measured for $\lambda$. 

\begin{figure}
 \centering
 \subfigure[]{\label{fig:histo_Re_01}\includegraphics[width=0.4\linewidth]{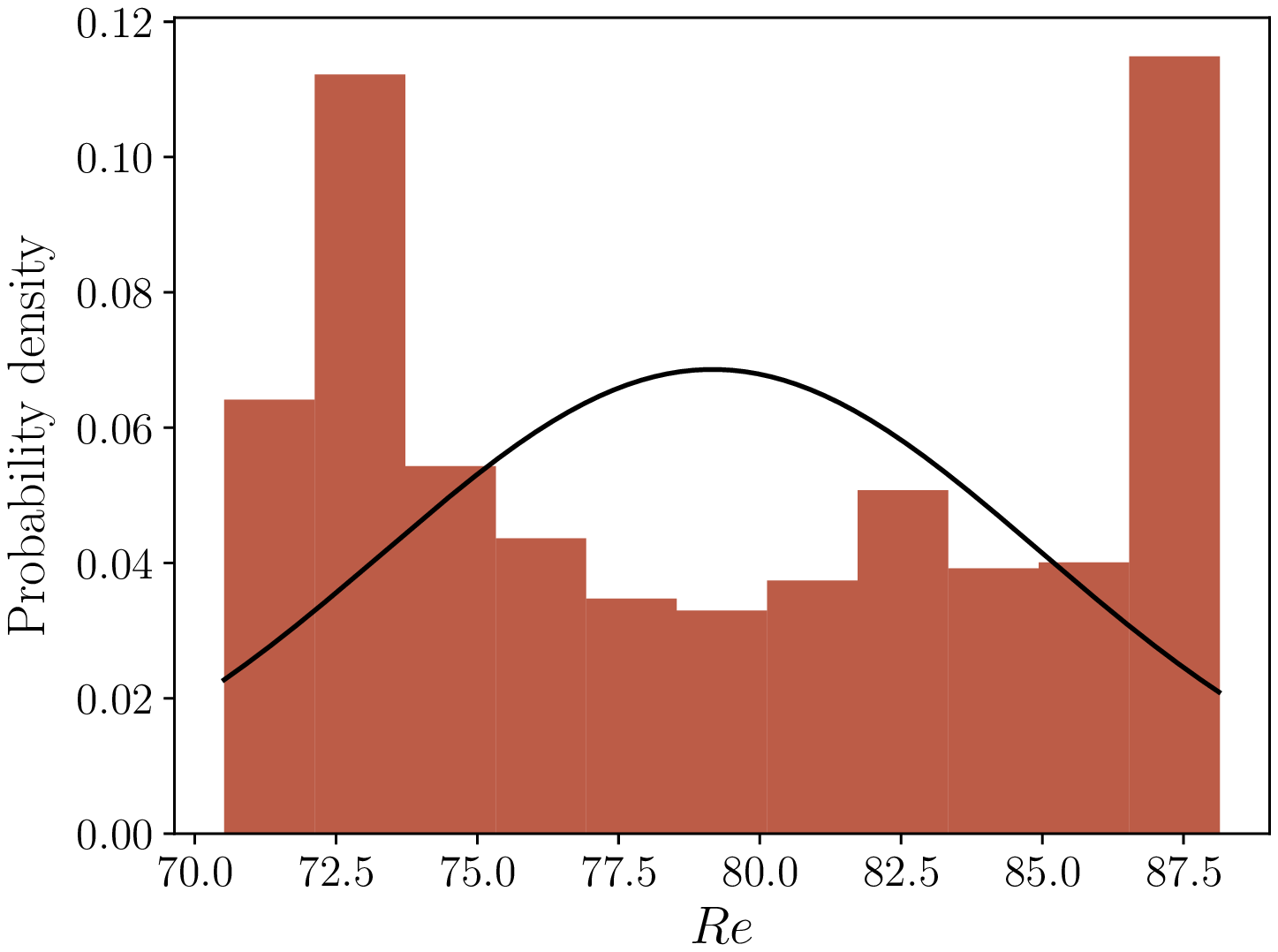}}
 \subfigure[]{\label{fig:histo_Re_06}\includegraphics[width=0.4\linewidth]{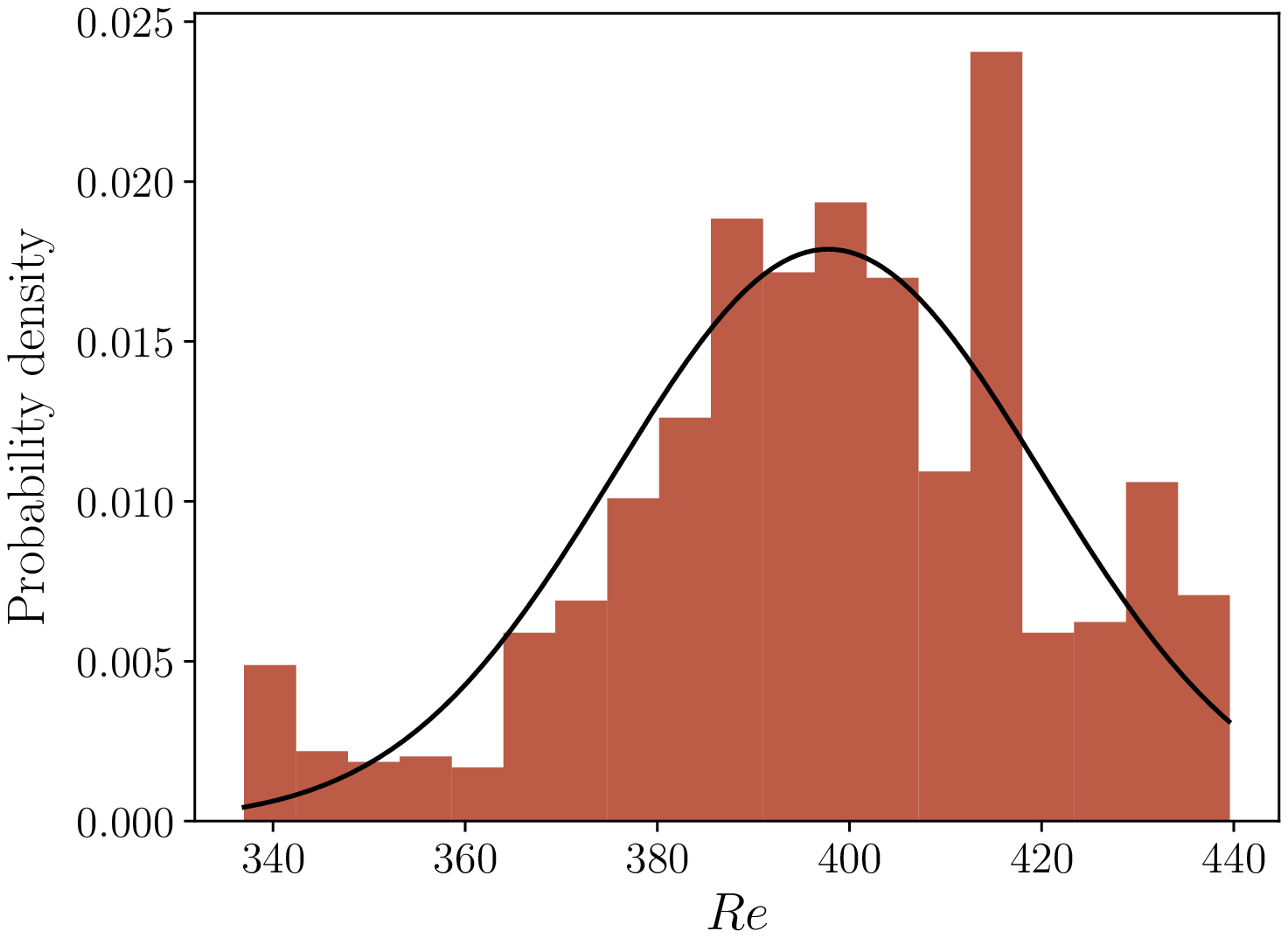}}
 \subfigure[]{\label{fig:histo_Re_10}\includegraphics[width=0.4\linewidth]{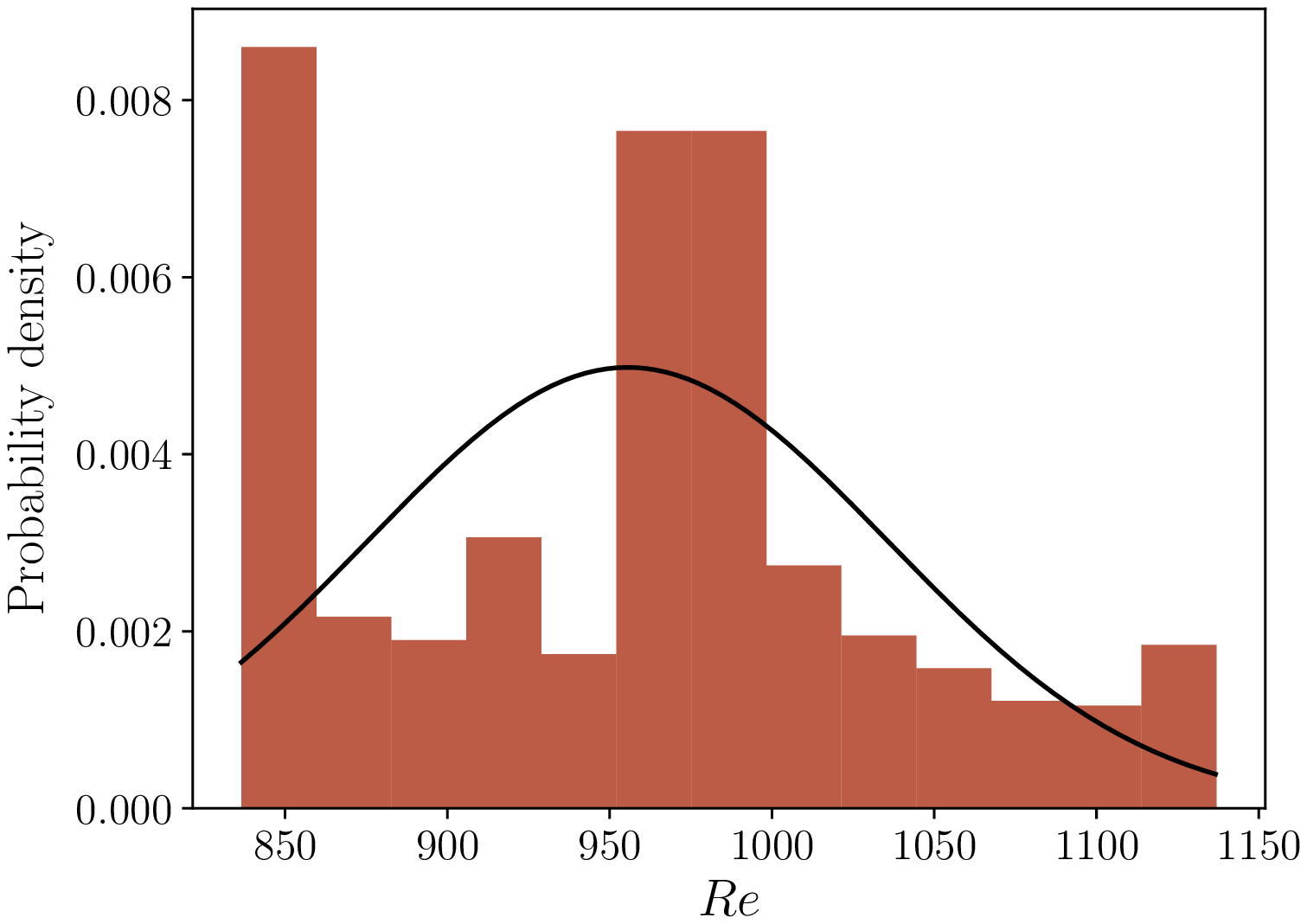}}
\caption{\label{fig:Rehistos} Normalized distributions of Re for three simulations: (a) $N=128$ and $\nu = 0.01$ , (b) $N=256$ and $\nu = 0.0018$, (c) $N=512$ and $\nu = 0.0008$. The black line represents the Gaussian function with mean and variance given by the distribution.}
\end{figure}

\begin{figure}
\centering
 \subfigure[]{\label{fig:histo_lyap_01}\includegraphics[width=0.4\linewidth]{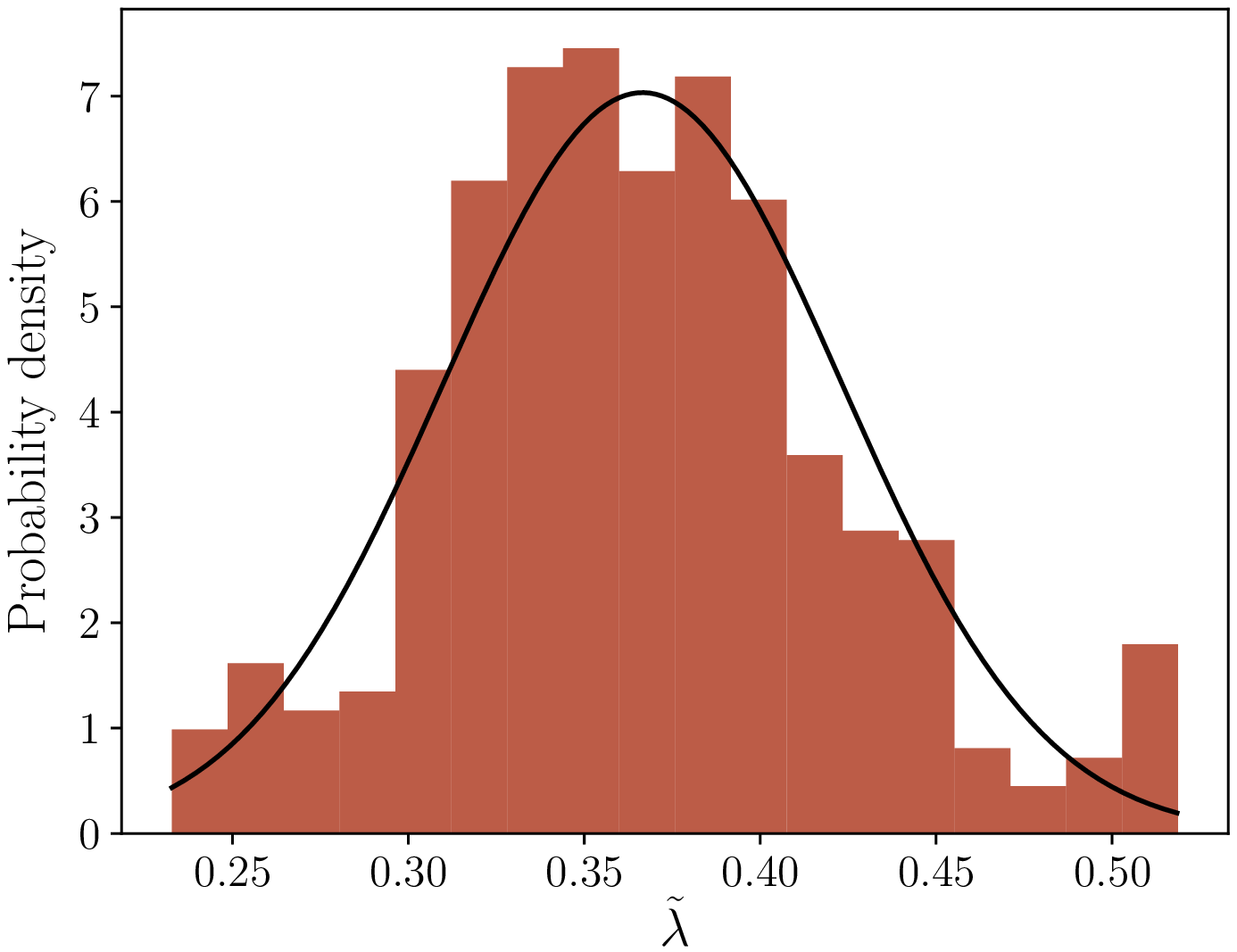}}
 \subfigure[]{\label{fig:histo_lyap_06}\includegraphics[width=0.4\linewidth]{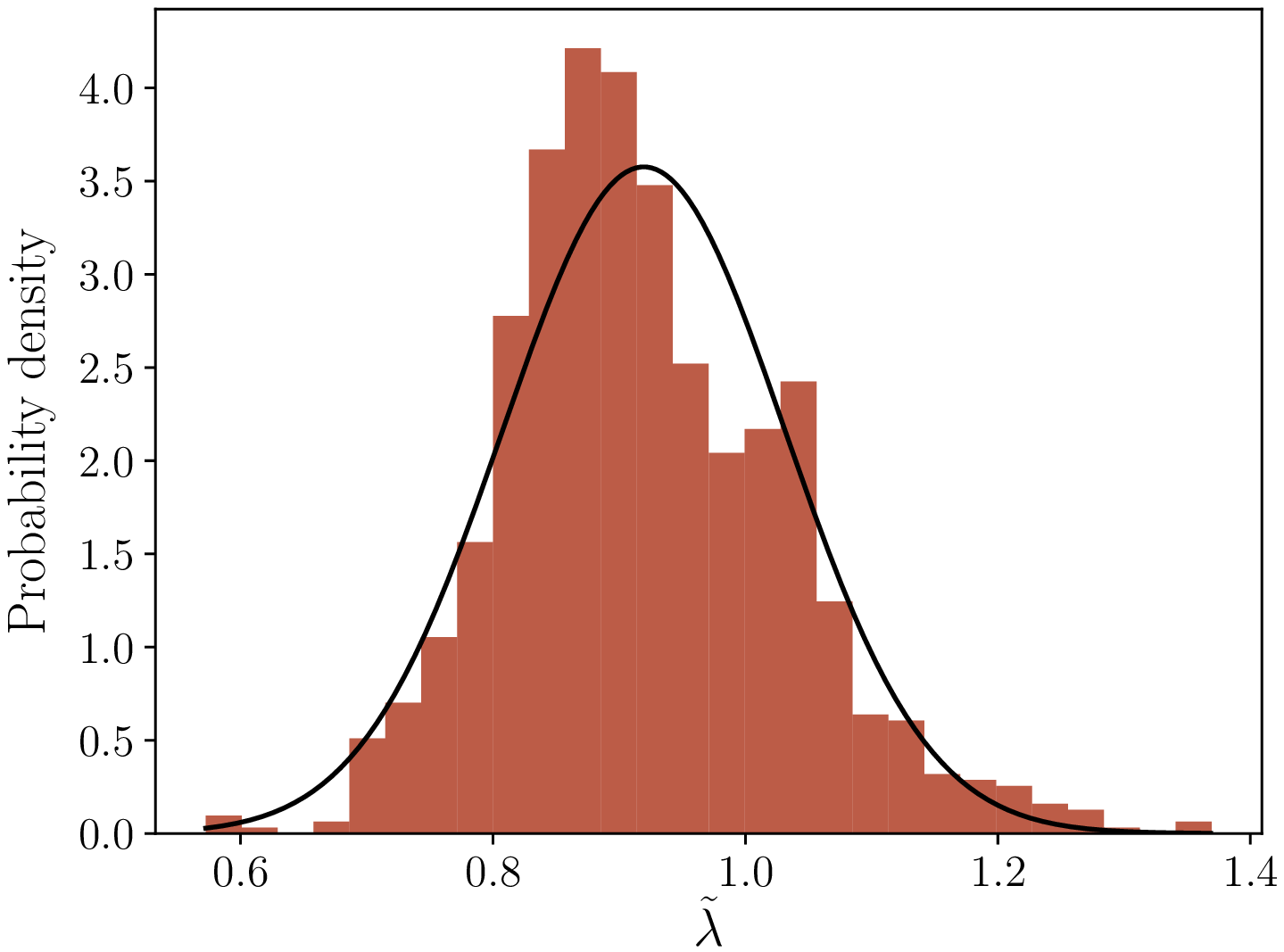}}
 \subfigure[]{\label{fig:histo_lyap_10}\includegraphics[width=0.4\linewidth]{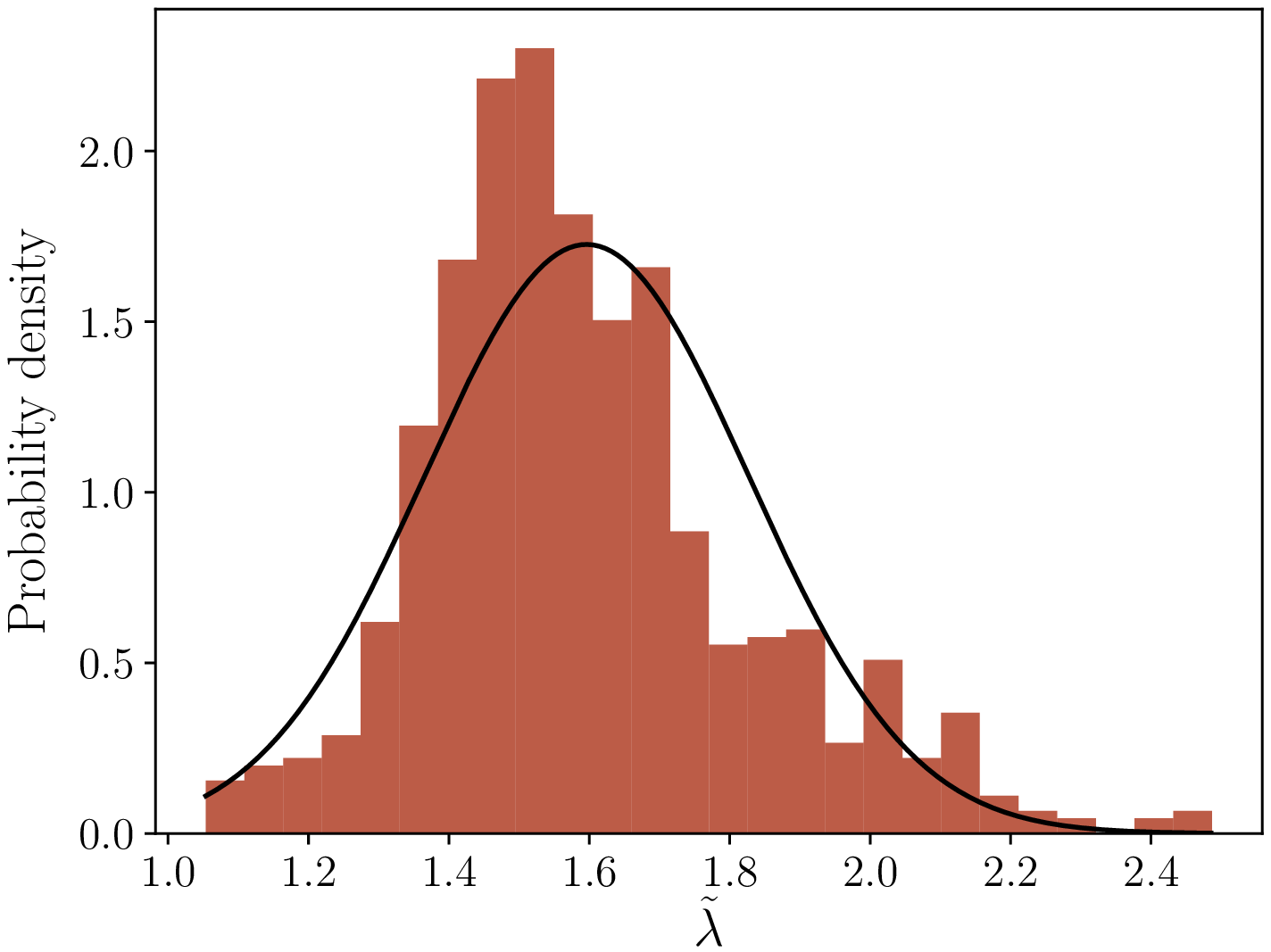}}
\caption{\label{fig:lyaphistos} Normalized distributions of $\Tilde{\lambda}$ for three simulations: (a) $N=128$ and $\nu = 0.01$, (b) $N=256$ and $\nu = 0.0018$, (c) $N=512$ and $\nu = 0.0008$. The black line represents the Gaussian function with mean and variance given by the distribution.}
\end{figure}

In Figure \ref{fig:Rehistos} we show the
typical distributions obtained for Re in the steady state,
measured for roughly 10-20 large eddy turnover times.
In Figure \ref{fig:lyaphistos} we show the
typical distributions obtained for $\Tilde{\lambda}$ in the steady state,
measured for the same run time as the equivalent distribution for the Reynolds number.

\begin{figure}
    \centering
    \includegraphics[width = 8.6cm]{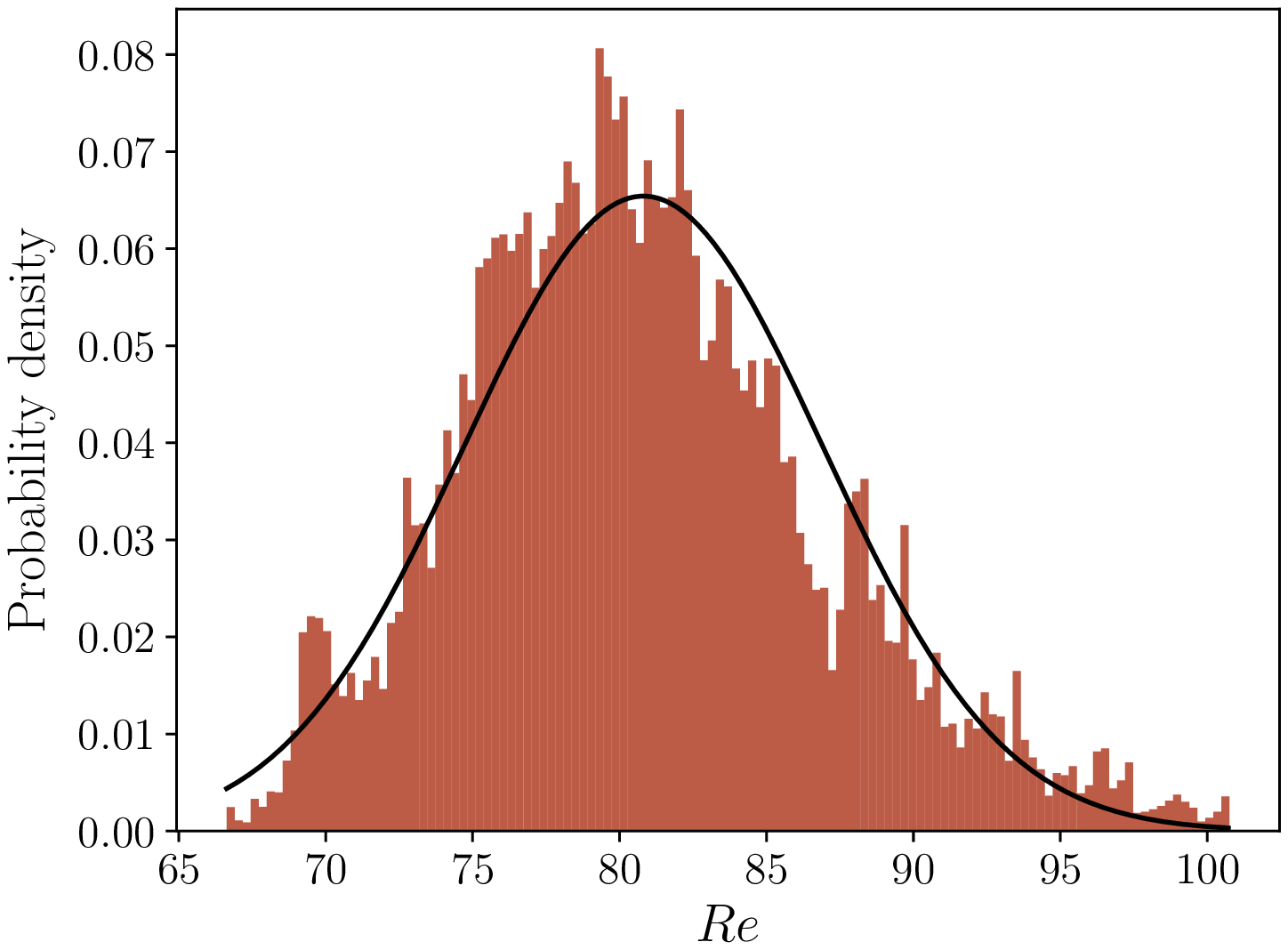}
    \caption{Histogram for the distribution of Re in a single run with measurement time of approximately $500\, T_0$. The solid black line represents the Gaussian function with mean and variance corresponding to the data sample}
    \label{fig:histo_Re_large}
\end{figure}

Surprisingly, most of the distributions of FTLEs are well approximated by a 
Gaussian whereas the Reynolds numbers histograms were poorly approximated
by a Gaussian distribution and even far from a bell shaped distribution in most cases.
For a much longer measurement time, the distributions
for Re and other large scale statistics approximate a Gaussian distribution more than
those in Figure \ref{fig:Rehistos}. The distribution of Figure \ref{fig:histo_Re_large} shows the distribution for a single run with a measurement time of approximately 500 large eddy turnover times, which corroborated that a Gaussian distribution is approximated for longer run times.
However, even after 10-20 large eddy turnover times, the Re statistics approximate
a Gaussian very poorly.
This result may be due to the fact that successive values of Re in a simulation are 
not random but strongly correlated, whereas for the Lyapunov exponent the correlation is weaker.

It is remarkable that given Equation (\ref{eq:ly_vs_Re^alph}), the distribution of Lyapunov 
exponents approximates to a Gaussian form when the Reynolds numbers does not even
come close to this. 
Apparently some source of randomization is introduced into the measurement, 
which could well be a consequence of the perturbation method, which occurs at a relatively high rate. This is one of the most important observations in this paper.

\begin{figure}
    \centering
    \includegraphics[width = 8.6cm]{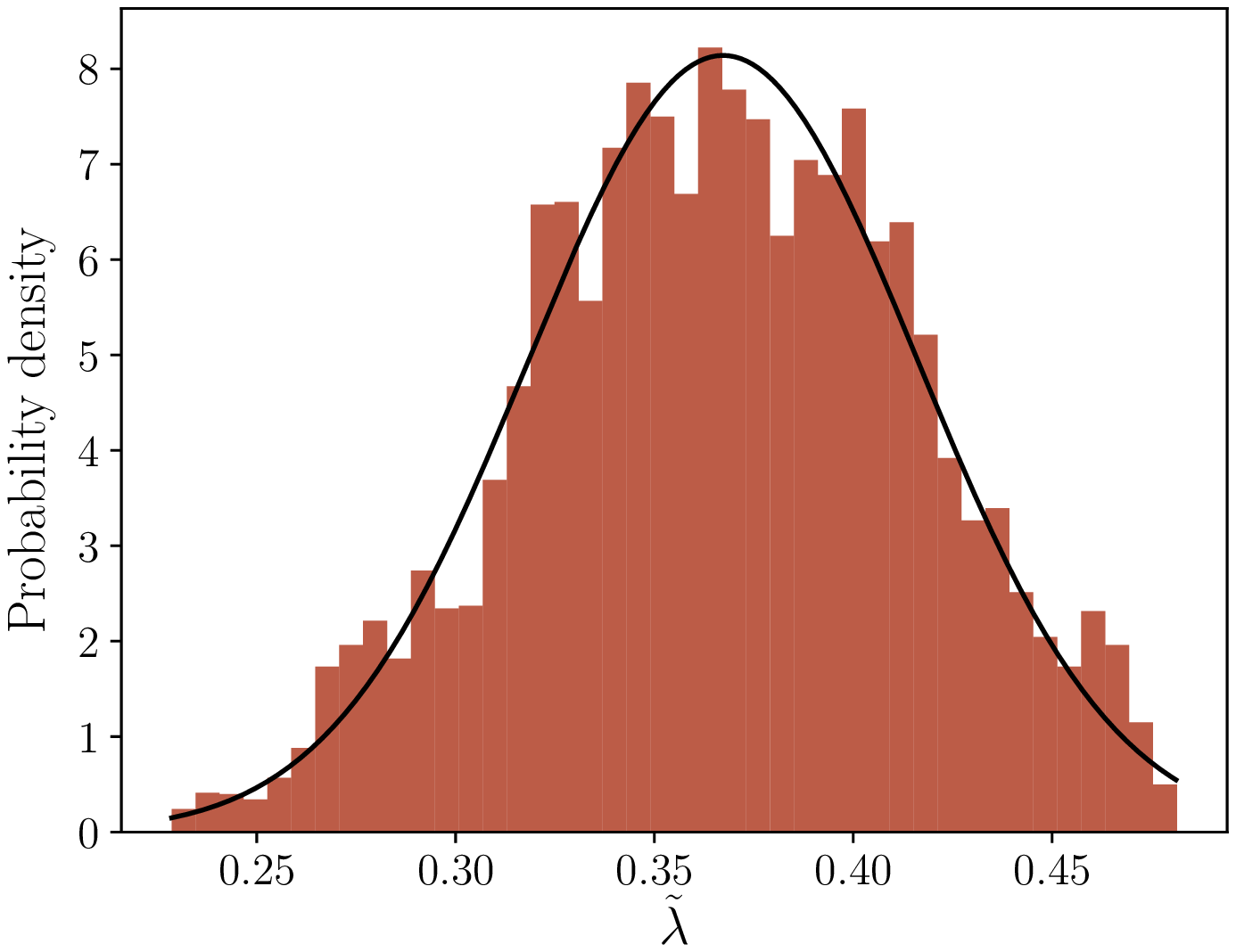}
    \caption{Histogram for the distribution of $\Tilde{\lambda}$ in a single run with $N = 128$ and $\nu = 0.01$, obtaining a large sample of $n = 11863$. The solid black line represents the Gaussian function with variance and mean given by the numerical data.}
    \label{fig:histo_large}
\end{figure}

In Figure \ref{fig:histo_large}, we present the distribution of $\lambda$ for a simulation with a run time of $200\, T_0$, obtaining 
$n = 11863$ values for $\Tilde{\lambda}$.
From this Figure, we can see that, 
even though FTLE distributions approximate to a Gaussian, 
it does not tend to an exact normal distribution for large $n$ and some noise is 
still observed.
The distribution in this Figure has skewness -0.12 and flatness 2.66, where
skewness is the third standardized moment of the probability
distribution and flatness is the fourth standardized moment.
For a Gaussian distribution we would expect skewness to be 0 and flatness to be 3.
However, for this simulation, the skewness of total energy, Re, and dissipation
is always positive and in the range 0.42-0.46. Flatness is slightly above 3
for these statistics.

We may measure the FTLE immediately after initializing the simulation, or wait
until after the system reaches a statistically steady state.
If the FTLE is measured immediately after initialization, there is a initial
transient until it reaches a statistically steady state, but this transient
is shorter than for the other large scale statistics such as total energy and Reynolds
number. This is also provides evidence that the correlation timescale for the Lyapunov exponent
is much shorter than for Reynolds number, energy and dissipation rate.

\label{se:results}
\subsection{Dependence of $\lambda$ and Re on lattice size}
\label{se:lattice}

We look at the dependence of the Lyapunov exponent $\lambda$ and its fluctuations $\sigma_{\lambda}$ on the lattice size. 
The Lyapunov exponent, Reynolds number and their fluctuations are measured for 
different runs keeping viscosity and dissipation rate fixed 
($\nu = 0.01$ and $\varepsilon \sim 0.1$), 
only varying the lattice size (i.e. $N = 32, 64, 128, 256, 512$) 
and random seed for the initial random field. 
It is important to mention that for each simulation, 
the mean Reynolds Number obtained are not always the same, 
ranging between $\sim 75-90$.

\begin{figure}
    \centering
    \includegraphics[width = 8.6cm]{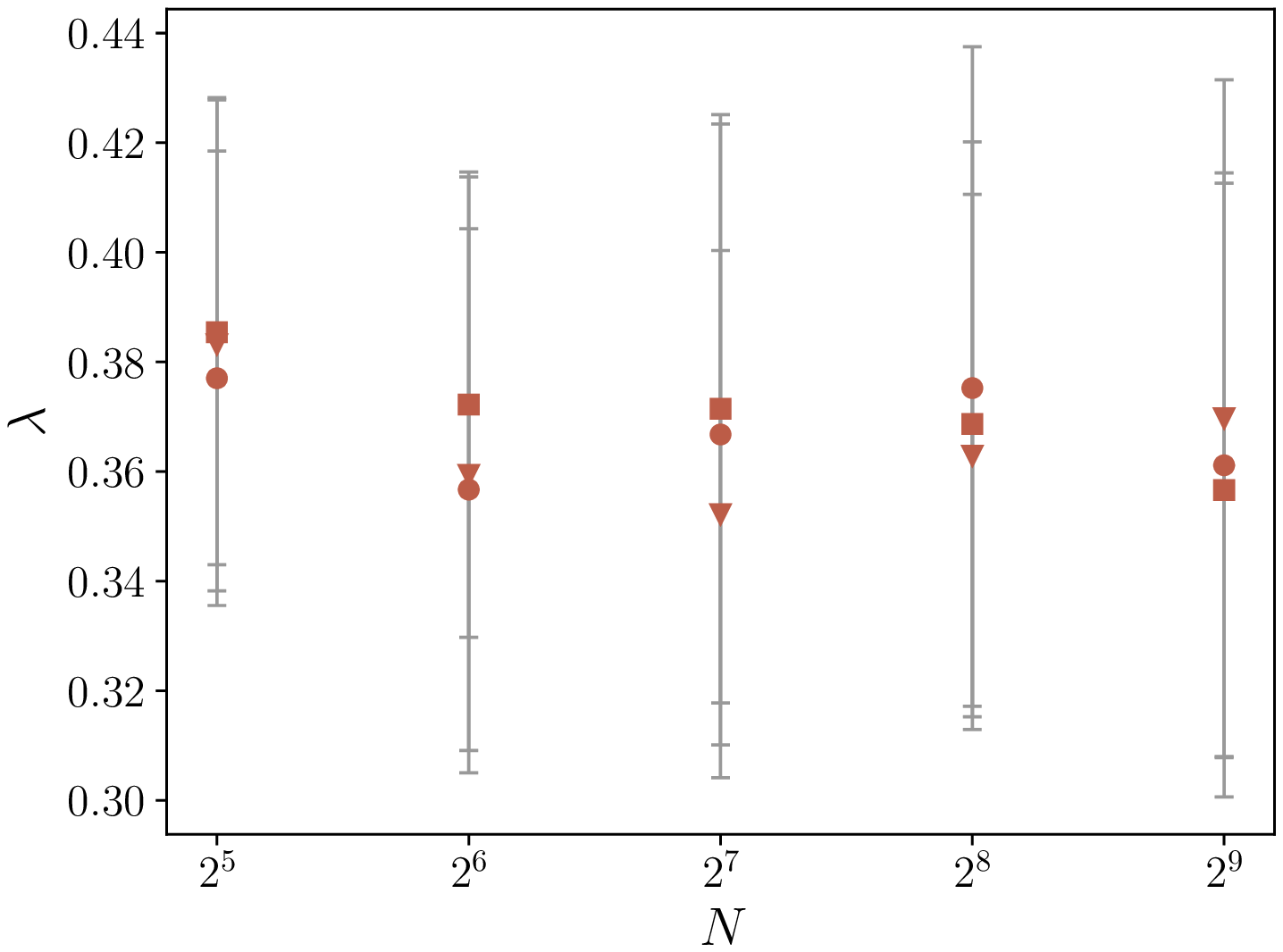}
    \caption{Dependence of $\lambda$ on lattice size $N$, different shapes represent
    different initial seeds.}
    \label{fig:ly_vs_N}
\end{figure}

\begin{figure}[h!]
    \centering
    \includegraphics[width = 8.6cm]{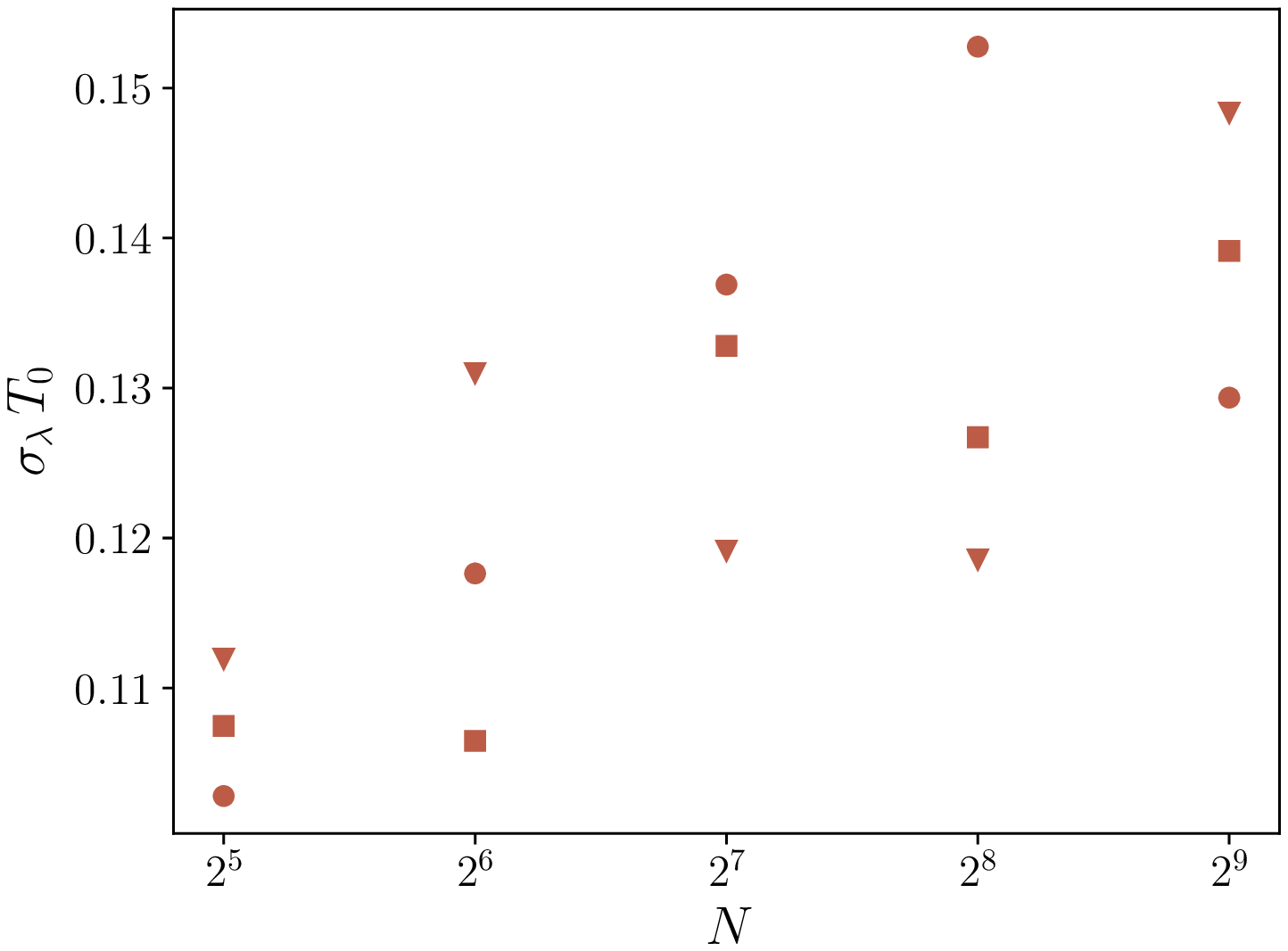}
    \caption{Dependence of Lyapunov exponent fluctuations $\sigma_{\lambda} T_0$ vs. $N$, different 
    shapes represent different seeds for the initial random field for each lattice size}
    \label{fig:DlyT_v_N}
\end{figure}

In Figure \ref{fig:ly_vs_N}, we show the dependence of $\lambda$ on lattice size N
for different initial seeds. As can be seen, the Figure shows no correlation or dependence
of $\lambda$ on lattice size. The mean values $\lambda$ varies little with $N$ compared to the fluctuations represented in the error bars. In Figure \ref{fig:DlyT_v_N}, we show the dependence of the fluctuations of the Lyapunov exponent $\sigma_{\lambda}$
on the lattice size.
A slight trend is observed in the Figure, although the variation is not significant compared to that given by the variation in Re in the following section and the uncertainty in the value of $\sigma_{\lambda}$ is not relevant given that this affects only the second significant figure in every measurement we perform.


\subsection{Dependence of $\lambda$ on Re}
\label{se:Re_dependence}

\begin{figure}
    \centering
    \includegraphics[width = 8.6cm]{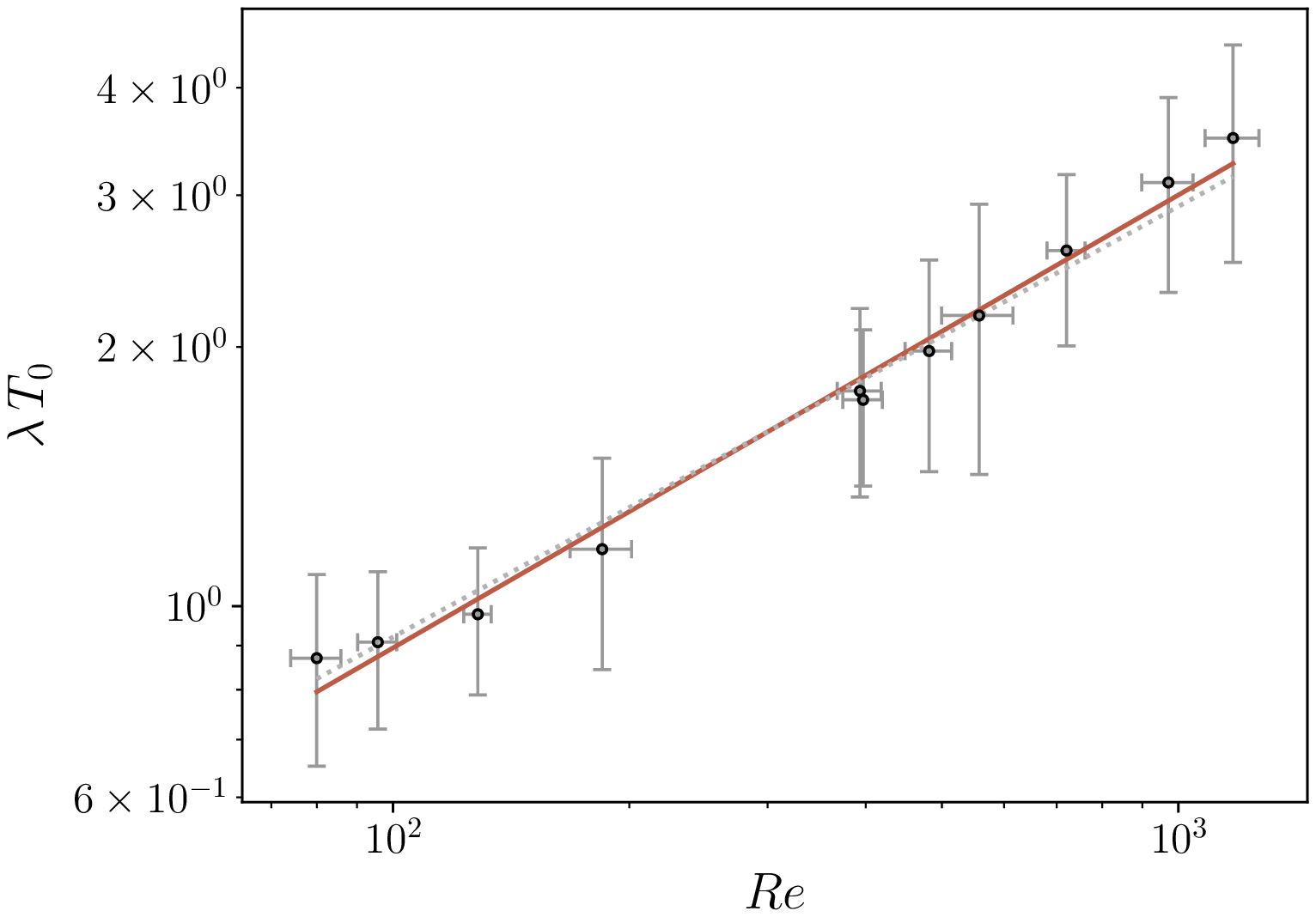}
    \caption{(Color online) $\lambda T_0$ vs. Re obtained for eleven simulations where $T_0 = L/U$. The linear fit is represented by a solid gray (red) line and Ruelle's prediction is represented by the dotted gray line.}
    \label{fig:lyT_v_R}
\end{figure}

After determining the variation of the fluctuations with the lattice size, we observe how $\lambda$ and its fluctuations 
vary depending on the Reynolds number. For this, eleven simulations were run varying viscosity in order to obtain different values of Re ranging between 80 and 1200. Table \ref{tab:NSE_sim} shows parameter values for all simulations. 
Once the values for $\lambda, \sigma_{\lambda}$,Re and $\sigma_{\text{Re}}$ are obtained,
a weighed least squared fit is performed as shown in Figure \ref{fig:lyT_v_R}
to Equation (\ref{eq:ly_vs_Re^alph}), obtaining a value for $\alpha = 0.53 \pm 0.02$ 
and $D = 0.079 \pm 0.007$. 
These results are in agreement with those previously found in \cite{Berera2018} 
in which the same forcing scheme is used but instead of FTLE, the direct method is used.
\begin{table}[]
    \centering
\begin{tabular}{cccccccccc}
\toprule
$N^3$ &   $\nu$ &  $\varepsilon$ &      Re &  $\lambda$ &  $\sigma_{\lambda}$ &   $T_0$ &  $T_{E0}$ &  $\tau$ &  $k_{\text{max}}\eta$ \\
\hline
 $128^3$ &  0.0100 &          0.096 &    80.0 &    0.359 &            0.057 &  2.43 &    5.17 &   0.323 &                  2.39 \\
 $128^3$ &  0.0080 &          0.097 &    96.0 &    0.396 &            0.057 &  2.29 &    5.15 &   0.287 &                  2.01 \\
 $128^3$ &  0.0060 &          0.097 &   128.0 &    0.445 &            0.064 &  2.20 &    5.43 &   0.249 &                  1.62 \\
 $128^3$ &  0.0040 &          0.101 &   185.0 &    0.573 &            0.099 &  2.03 &    5.39 &   0.199 &                  1.18 \\
 $256^3$ &  0.0020 &          0.098 &   393.0 &    0.871 &            0.153 &  2.04 &    5.89 &   0.143 &                  1.44 \\
 $256^3$ &  0.0018 &          0.099 &   397.0 &    0.899 &            0.123 &  1.93 &    5.60 &   0.135 &                  1.32 \\
 $256^3$ &  0.0016 &          0.098 &   482.0 &    0.991 &            0.194 &  2.00 &    5.90 &   0.128 &                  1.21 \\
 $256^3$ &  0.0014 &          0.099 &   558.0 &    1.096 &            0.256 &  1.98 &    5.97 &   0.119 &                  1.10 \\
 $512^3$ &  0.0010 &          0.100 &   721.0 &    1.357 &            0.218 &  1.91 &    5.66 &   0.100 &                  1.70 \\
 $512^3$ &  0.0008 &          0.099 &   971.0 &    1.582 &            0.270 &  1.96 &    5.97 &   0.090 &                  1.44 \\
 $512^3$ &  0.0006 &          0.100 &  1174.0 &    1.877 &            0.367 &  1.86 &    5.67 &   0.077 &                  1.16 \\

\hline
\end{tabular}
    \caption{Simulation parameters for 11 simulations using incompressible NSE. N is the lattice size, $\nu$ is the kinematic viscosity, $\epsilon$ is the dissipation rate, Re is the Reynolds number, $\lambda$ is the maximal Lyapunov exponent and $\sigma_{\lambda}$ its standard deviation, $T_0=L/U$ and $T_{E0} = E/\varepsilon$ are two different definitions for the large eddy turnover time, $\tau$ is the Kolmogorov microscale time, $k_{\text{max}}\approx N/3 -1$ is the maximum resolved wavenumber, $\eta = (\nu^3/\varepsilon)^{1/4}$ is the Kolmogorov microscale length and $k_{\text{max}}\eta$ is the resolution.}
    \label{tab:NSE_sim}
\end{table}

The novel analysis here is done on the fluctuations (deterministic fluctuations as discussed in the Introduction). 
First, the fluctuations in Re, $T_0$ and $\lambda$ are studied. 
Then, using the relation in Eq.(\ref{eq:ly_vs_Re^alph}), the fluctuations of the quantities on the r.h.s are propagated to estimate the expected level of fluctuations of the Lyapunov exponent on the l.h.s, thus corroborating if the level of fluctuations is consistent on both sides. It is observed that the level of fluctuations in $\lambda$ is larger than that expected from Eq.(\ref{eq:ly_vs_Re^alph}). This might indicate that there are other effects on the system introducing these variations, although it is important to recall that Eq.(\ref{eq:ly_vs_Re^alph}) is not to be taken as a fundamental relation, since it is derived from dimensional considerations. According to the theory, FTLEs have some dependence on the initial finite perturbation and the finite time in which it is measured, so the fluctuations can have origin not only in the fluctuations of the steady state but also by the successive perturbations introduced by the FTLE method and the steptime $\Delta t$.

\begin{figure}[h!]
\centering\includegraphics[width=8.6cm]{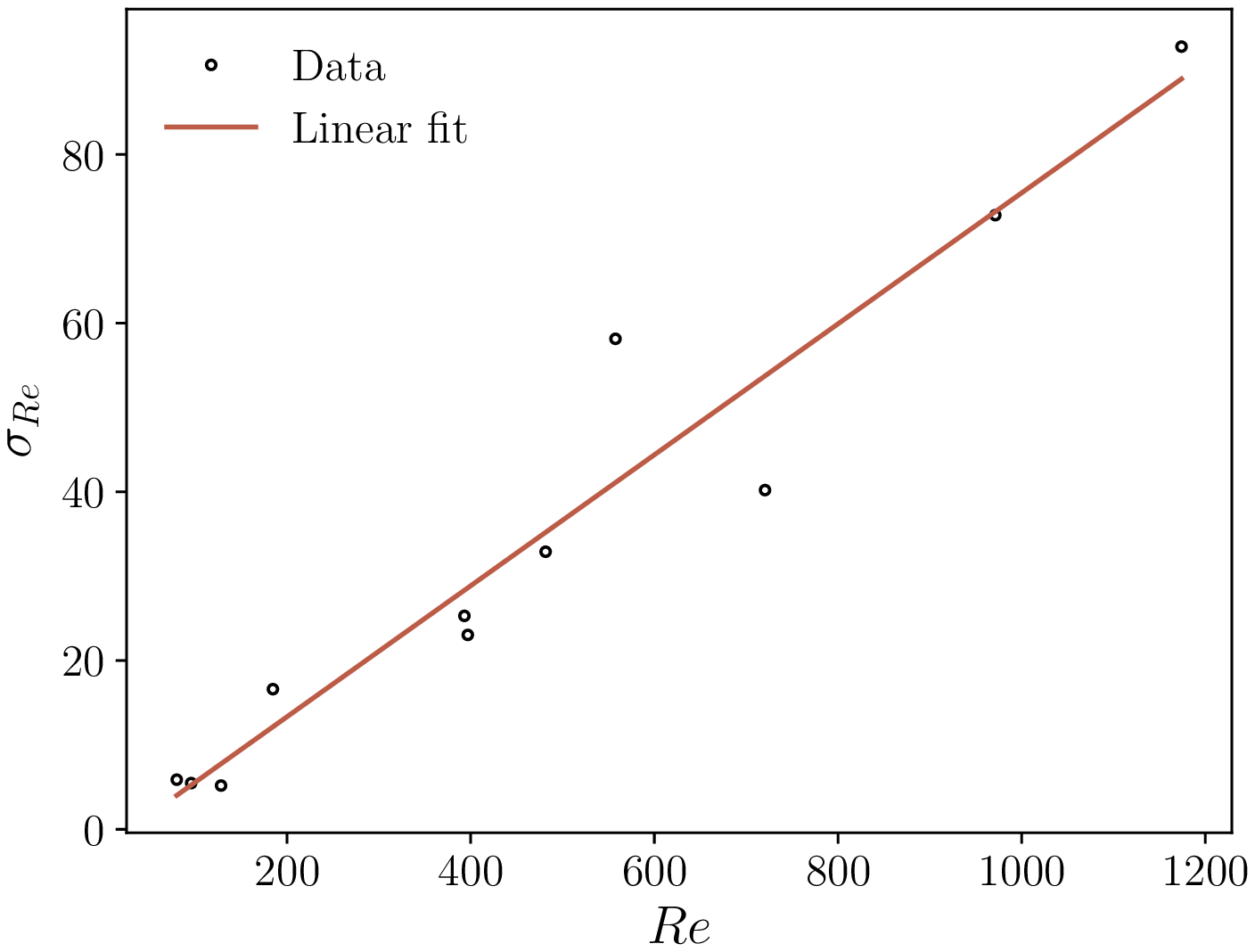}
\caption{$\sigma_{\text{Re}}$  vs. Re for statistically steady state of the flow and linear fit for slope $c_1 = 0.08 \pm 0.01$}
\label{fig:DRevsRe}
\end{figure}

Figure \ref{fig:DRevsRe} plots the fluctuations in Re, $\sigma_{\text{Re}}$, 
against their mean values.
It shows that there is a linear dependence in the observed range of Re.
The following relation is found, 
$\sigma_{\text{Re}} = c_1 \text{Re}$, with $c_1 = 0.08 \pm 0.01$. Similarly for the relation between $\lambda$ and $\sigma_{\lambda}$, a linear dependence is observed and
the following relation is measured, 
$\sigma_{\lambda} = c_2 \lambda$, with $c_2 = 0.20 \pm 0.02$. In our simulations,
the values for $T_0$ and $\sigma_{T_0}$ are approximately constant, with $\sigma_{T_0}/T_0 \sim 0.05$.

To corroborate that these fluctuations are 
not consistent with Equation (\ref{eq:ly_vs_Re^alph}), we propagate $\sigma_{T_0}$ and $\sigma_{\text{Re}}$ and using the measured values we compare the result to the $\sigma_{\lambda}$ obtained using the FTLE procedure. All fluctuations are deterministic and are measured in the steady state. The following relation is obtained
\begin{align}
\sigma_{\lambda} \approx   \left\lvert\frac{\partial \lambda}{\partial Re} \right\rvert \sigma_{Re} + \left\lvert\frac{\partial \lambda}{\partial T_0} \right\rvert \sigma_{T_0}  
\end{align}
Here, we consider $D$ and $\alpha$ as universal constants with no fluctuations associated to them. Their uncertainties do not stem from deterministic fluctuations as for Re and $T_0$, instead they have origin in the linear regression performed to estimate their values. This distinction is highly important for this analysis. To compare fluctuations, it is advantageous to consider the standard deviation of $\lambda$ relative to its mean value, so we expect
\begin{align*}
\frac{\sigma_{\lambda}}{\lambda} \approx \alpha \, \frac{\sigma_{Re}}{Re} + \frac{\sigma_{T_0}}{T_0} =  0.09 \  .
\end{align*}
Comparing this expected value to the measured value $\frac{\sigma_{\lambda}}{\lambda} = 0.2$, we find that the level of fluctuation we expect from Eq.(\ref{eq:ly_vs_Re^alph}) is inconsistent with the one measured. Such fluctuations appear to have origin not only in the fluctuations of $T_0$ and Re. There are many possible explanations for the high level of fluctuations in $\lambda$. The perturbation in the FTLE procedure or the forcing could affect the value of $\sigma_{\lambda}$. Another possible cause is that even for infinitesimal perturbations, turbulent flows are not characterized by a unique maximal Lyapunov exponent, contrary to what the theory indicates. The reason might well be a combination of all those previously mentioned. The important fact is that $\sigma_{\lambda}/\lambda$ fluctuations are significant and they cannot be simply estimated by measuring the values of Re, $\sigma_{\text{Re}}, T_0$ and $\sigma_{T_0}$.  

\begin{figure}[h!]
    \centering
    \includegraphics[width = 8.6cm]{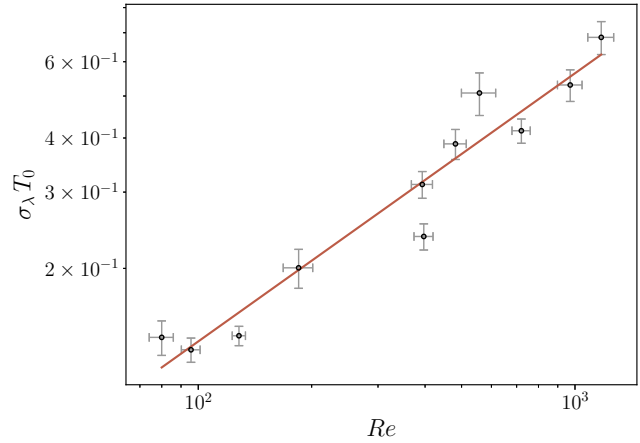}
    \caption {Data and linear fit to determine power law $\sigma_{\lambda} T_0 \sim \text{Re}^{\gamma}$ }
    \label{fig:DlyT_v_R}
\end{figure}

The relationship between fluctuations of the Lyapunov exponent and the mean Reynolds number is
also measured.
Figure \ref{fig:DlyT_v_R} shows a plot of $\sigma_{\lambda}$ against Re 
with a fit to the scaling relation $\sigma_{\lambda} T_0 \propto \text{Re}^{\gamma}$. 
The value found for our simulations is $\gamma =0.62 \pm 0.05$.

\subsection{Effect of definition of $T_0$ on $\alpha$}
\label{se:def_of_T_0}

We note that previous results which have analyzed the relationship in
Equation (\ref{eq:ly_vs_Re^alph}) have found disagreement in the value of
$\alpha$ which they produce.
For example, previous results using the direct method for DNS have given
an $\alpha = 0.53 \pm 0.03$ \cite{Berera2018}, which agrees with the finding using an FTLE
method as done here.
However, other FTLE DNS results have found $0.64\pm0.05$ \cite{Boffetta2017}.
This section offers a resolution to this discrepancy in the results.

One of the main features of turbulence is that it involves many characteristic scales. 
Furthermore, in homogeneous and isotropic turbulence, the absence of boundary conditions sets 
an extra complication in determining the large scale quantities. Using different definitions of the large length and time scales can result in different
analyses.
The key difference between the prior works was that
these used different choices for the definition of $L$ and $T_0$. 
In \cite{Berera2018}, $L$ is the integral length scale defined in 
section \ref{S:HD} and $T_0 = L/U$, whereas in \cite{Boffetta2017}, 
this value was defined differently, being $T_{E0} = E/\varepsilon$ 
and then $L_E=UT_{E0}$. 
The forcing used in \cite{Berera2018} was the same used in this work
and the same as used in \cite{Boffetta2017}.  

The difference between these definitions deserves some attention.  
The definition of the integral length scale $L$ stems from the correlation function $R_{ij}(r) = \langle u_i(\bm{x})u_j(\bm{x}+\bm{r}) \rangle$ 
and can be associated with the size of the largest eddies. 
Hence, the first definition, given by the integral length scale, 
can be understood as the average time that it takes for a large eddy 
initially occupying a given region is space to move to a different region 
which is completely uncorrelated with the initial one. The second definition $T_{E0}$ takes the total energy $E$ 
and the dissipation rate $\varepsilon$, 
which is associated with the characteristic time in which the largest eddies transfer energy to smaller eddies in the energy cascade.  

Taylor in 1935 proposed the following relation for a HIT flow in a steady state, 
using dimensional analysis \cite{Taylor35},
\begin{equation}
\varepsilon = C_{\varepsilon} \frac{U^3}{L} \quad ,
\end{equation} 
where $C_{\varepsilon}$ is known as the dimensionless dissipation rate. 
Evidence has shown that this coefficient tends to a constant value 
$C_{\varepsilon,\infty} \approx 0.5$ for high Re \cite{doering1994variational,sreenivasan1998update,gotoh2002velocity}.

At this point we should note that if we compare the definitions of $T_0$ and $T_{E0}$ in the large Reynolds numbers limit where $C_{\varepsilon} \approx 0.5$ and using that $E=3U^2/2$, these two definitions differ only by a constant factor, being $T_{E0} = 3 T_0$. In such case, there should be no discrepancy when using these different definitions to test Ruelle's relation. Nevertheless, if we consider that the dimensionless dissipation has some Reynolds number dependence, then both definitions will have an impact in the scaling relation between $\lambda$ and Re given by Eq.(\ref{eq:ly_vs_Re^alph}). 

We can improve the accuracy of this analysis by considering a useful expression for the dimensionless dissipation rate derived by McComb \textit{et al} \cite{Mccomb2015}. The following expression is derived using the von K\'arm\'an-Howarth 
equation and performing an asymptotic expansion in the second and third-order structure functions in powers of the inverse Reynolds number, obtaining
\begin{equation}
\label{eq:Ceps}
    C_{\varepsilon}(Re) = C_{\varepsilon,\infty} + \frac{C}{Re}  + \mathcal{O}\left(\frac{1}{Re^2}\right) \quad , 
\end{equation}
where $C$ is a constant that does not depend on Re. 
Also in \cite{Mccomb2015}, these constants are computed using DNS. 
The numerical values obtained are 
 $C_{\varepsilon,\infty} = 0.486\pm0.006 $ and $C = 18.9 \pm 1.3$,
which are in agreement with other values in literature \cite{sreenivasan1998update}. Our simulations show that the relation of $C_{\varepsilon}(\text{Re})$ as a function of Re is in 
strong agreement with the results of \cite{Mccomb2015}.

\begin{figure}
    \centering
    \includegraphics[width = 8.6cm]{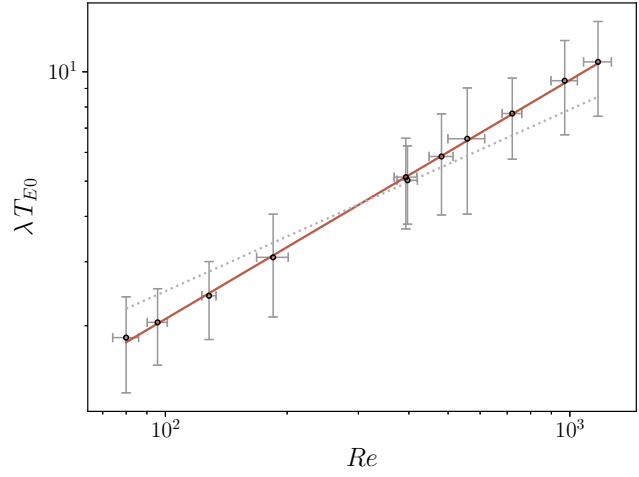}
    \caption{(Color online) $\lambda$ vs Re using the definition of $T_{E0} = E/\varepsilon$. The linear fit is indicated by a solid gray (red) line whereas Ruelle's prediction is denoted by the dotted line.}
    \label{fig:lyT_v_R_dd2}
\end{figure}

Considering this result, we see that the relation between $T_{E0}$ and $T_0$ 
carries some extra dependence on the Reynolds number, 
\begin{align}
    T_{E0} &= \frac{E}{\varepsilon} = \frac{3L}{2C_{\varepsilon}U} =  \frac{3}{2 C_{\varepsilon}(Re)} T_0 \quad .  
    \label{eq:T0_TE0}
\end{align}
The scaling relation in Equation (\ref{eq:ly_vs_Re^alph}) is re-tested using the definition of $T_{E0}$. 
We expect that when we use the definition $T_{E0}$ instead of $T_0$, a different exponent $\alpha_{E}$ is found instead of $\alpha$. So
\begin{equation}
\lambda T_{E0} = D_E Re^{\alpha_E} \quad .
\label{eq:ly_vs_Re_dd}
\end{equation}
Figure \ref{fig:lyT_v_R_dd2} shows the data for $\lambda$ as a function of Re
using the definition $T_{E0}$.
Using a linear fit, we obtain the values for Equation (\ref{eq:ly_vs_Re_dd})
of $\alpha_E = 0.658 \pm 0.006$ and $D_E = 0.101 \pm 0.001$. These results are in agreement with those found in \citep{Boffetta2017}.

\begin{figure}[h!]
\centering
\includegraphics[width = 8.6cm]{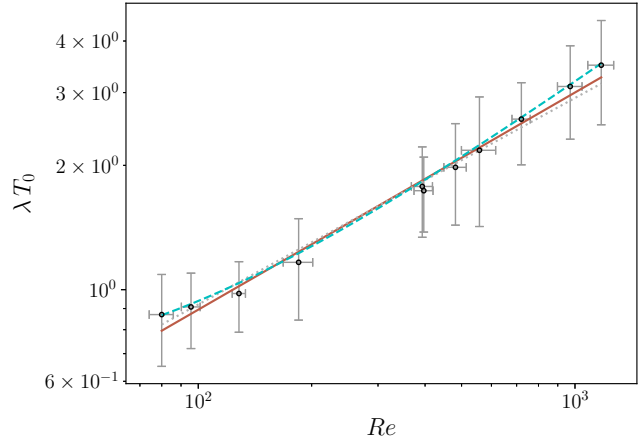}
\caption{(Color online) Data obtained for $\lambda$ vs $T_0$, the straight solid gray (red) line represent the linear fit, the straight grey dotted line represents Ruelle's prediction and the curved dashed light gray (cyan) line represents the prediction given by $f($Re$)$.}
\label{fig:Boff_Ho}
\end{figure}

Since the data in Figure \ref{fig:lyT_v_R_dd2} seems to be more correlated than that in Figure \ref{fig:lyT_v_R},
we can take Eq.(\ref{eq:ly_vs_Re_dd}) as the base equation and see how it
would affect Eq.(\ref{eq:ly_vs_Re^alph}).
In this way we attempt to compare the two different definitions $T_0$ and $T_{E0}$. 
Starting from Eq.(\ref{eq:ly_vs_Re_dd}), the following relation is obtained 
\begin{align}
        \lambda \overbracket{T_0 \frac{3}{2 \, C_{\varepsilon}}}^{T_{E0}}&= D_E Re^{\alpha_E} \, , \nonumber\\
        \lambda T_0 &= f(Re) \equiv \frac{2 D_E}{3} Re^{\alpha_{E}} C_{\varepsilon}(Re) \, .
        \label{eq:f(Re)}
\end{align}
This relation sets an extra dependence on the Reynolds number that is not present in 
the original Ruelle's relation in Eq.(\ref{eq:ly_vs_Re^alph}). Figure \ref{fig:Boff_Ho} shows the comparison between the data obtained, 
Ruelle's prediction, and the prediction given by $f($Re$)$ in 
Eq.(\ref{eq:f(Re)}) by using 
the two different definitions $T_0$ and $T_{E0}$.
It can be seen that the slight curved deviation from the line is well predicted by $f($Re$)$.

We can also predict the expected value of $\alpha$ when considering the extra Re dependence in Eq.(\ref{eq:Ceps}) and given the measured value for $\alpha_E$ and see if it is consistent with the one found in the previous section.
To compare both exponents, we consider equations (\ref{eq:ly_vs_Re^alph}) and (\ref{eq:f(Re)}). Then we take the following derivative to determine the scaling relation 
\begin{align}
    \alpha &= \frac{d \ln(\lambda T_0)}{d \ln Re} = \alpha_E + Re \frac{d \ln\left(C(Re)\right)}{d Re}  \quad ,\nonumber \\
    \alpha &= \alpha(Re) = \alpha_E - \frac{1}{1 + \frac{C_{\varepsilon,\infty}}{C}Re} \quad . 
\end{align}
As expected from Figure (\ref{fig:Boff_Ho}), the scaling is not a constant since it has a slight dependence on Re. To compare with the expected value of $\alpha$ previously obtained, we first evaluate $\alpha($Re$)$ using the same values of Re measured in the simulations. Then we take the average, obtaining
\begin{equation}
    \langle \alpha \rangle_{\text{exp}} = \alpha_E - \left\langle  \frac{1}{1 + \frac{C_{\varepsilon,\infty}}{C}Re} \right\rangle \approx 0.52 \quad ,
\end{equation}
which is within one standard deviation of the measured value of $\alpha$. This shows that both procedures are consistent. Although the data in Figure \ref{fig:lyT_v_R_dd2} is more linearly correlated. This resolves the apparent disagreement between results in \cite{Berera2018} and \cite{Boffetta2017}, and is one of the main results in this paper.

We also measure the scaling relation $\sigma_{\lambda} T_{E0} \propto \text{Re}^{\gamma_E}$
and compare it to that found in \cite{Boffetta2017}.
In that work, the authors claim that the uncertainty for this value is large but obtain $\gamma_* \sim 1.2$.
In our simulations, we find a smaller value $\gamma_E = 0.75 \pm 0.05$,
which is closer to the value of $0.62 \pm 0.05$ for $\gamma$ found previously 
using the integral length scale to define $T_0$.

\subsection{FTLE steptime dependence of $\lambda$}
\label{se:Steptime_dependence}

When the FTLE method is used, 
it is necessary to set certain parameters such as 
the magnitude of the perturbation or the steptime $\Delta t$ between successive perturbations. 
It is reasonable to suppose that for large values of the steptime the results are more stable since the maximal Lyapunov exponent is defined for $t \rightarrow \infty$. However, in the case of FTLEs some dependence on the initial perturbation is expected, hence, the drawback is that for finite times, i.e. $\Delta t << 2T_{E0}$, the growth might not reach stability and thus the estimate of $\lambda$ and $\sigma_{\lambda}$ might be inaccurate for reasons we give in Section \ref{se:timescales_HIT}.
On the other hand, having shorter steptime would be an advantage in the sense that it allows one to obtain a larger sample of Lyapunov exponents in a significantly shorter simulation time.

\begin{figure}[h!]
    \centering
    \includegraphics[width = 8.6cm]{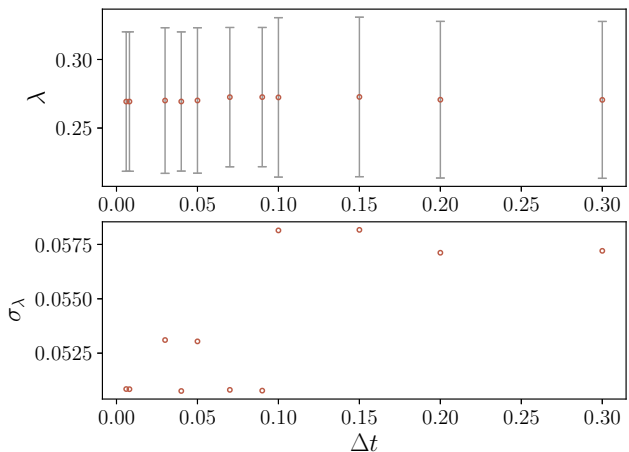}
    \caption{Dependence of $\lambda$ and $\sigma_{\lambda}$ on the steptime $\Delta t$.}
    \label{fig:Lambda_and_error_vs_steptime}
\end{figure}

Figure \ref{fig:Lambda_and_error_vs_steptime} shows the result of varying the steptime
$\Delta t$ on the Lyapunov exponent and its fluctuations.
The simulation has Re $= 50$.
As can be seen in the Figure, the values of $\lambda$ and $\sigma_{\lambda}$ are
remarkably stable even for very short steptimes using the FTLE method.
Indeed, at the extreme lowest value, 
the steptime is so short that it is only 5-10 simulation timesteps.
At the other extreme, the direct method, previously described, is akin to having
a very high steptime such that, within one steptime, the system becomes decorrelated.
Even using the direct method, the mean values of $\lambda$ are similar to the ones
found using the FTLE method.
These results suggest that the steptime is of very little importance to the measured
value of $\lambda$, and so it can be measured with a very low steptime which helps to reduce the computational cost. This finding can be useful in certain applications that use the Lyapunov exponent to characterize turbulent flows \cite{nastac2017lyapunov}, especially if we find a way to relate the Lyapunov exponent to other quantities that characterize the flow such as Reynolds number, energy or dissipation rate. This is the main finding in this paper and together with the fast stability described in Section \ref{se:timescales_HIT}. This means that the Lyapunov exponent is a robust measure in these type of simulations.

\subsection{Decaying turbulence}
\label{se:decaying}

We analyze the chaotic behaviour in a decaying turbulent flow. The procedure consists in evolving an initial field until it enters a power law decay phase. After that, a copy of the field is created and perturbed infinitesimally. Then, we measure the evolution of the energy difference in time $E_d(t)$ and compare it with predictions. Using the relation in Eq.(\ref{eq:ly_vs_Re^alph}) for steady states, we are able to predict the evolution of $E_d(t)$ by adapting it to a decaying flow. For this we consider the difference field at a given simulation time $t_i$ as $\lvert \delta \bm{u}\left( t_i\right)\rvert\equiv \delta u_{i}$. For each simulation time, we assume that the difference grows exponentially. Thus we have

\begin{equation}
    \delta u_{i+1} \approx \delta u_{i} \exp\left(\lambda_{i+1} \delta t\right)\frac{\lvert\bm{u}_{i+1}\rvert}{\lvert\bm{u_i}\rvert} \quad ,
\end{equation}
where $\delta t \equiv t_{i+1} - t_i$ is the timestep of the simulation. We multiply by the factor $\lvert\bm{u}_{i+1}\rvert/\lvert\bm{u_i}\rvert$ to account for the decay in the magnitude of $\left\lvert\bm{u}\right\rvert$, that we assume uniform. It is important to remark that this approximation also assumes that the Lyapunov exponent vary slowly in time, thus $\left(\lambda_{i+1}+\lambda_i\right)/2 \approx \lambda_{i+1}$.

Continuing for $i+2$, we obtain
\begin{equation}
    \delta u_{i+2} = \delta u_{i+1} \exp\left(\lambda_{i+1} \delta t + \lambda_{i+2} \delta t\right)\frac{\lvert\bm{u}_{i+2}\rvert}{\lvert\bm{u_i}\rvert} \quad ,
\end{equation}
and so, in general 

\begin{equation}
\delta u_f = \delta u_i \exp\left(\int_i^f dt \lambda(t)\right) \frac{\lvert\bm{u}_{f}\rvert}{\lvert\bm{u_i}\rvert} \quad ,
    \label{eq:decaying_predict}
\end{equation}
where we replace the infinite sum for the integral over time, and $\lambda (t)$ is a function of time. 

We use this prediction to discriminate between the different functional dependencies of $\lambda$ given in Eqs. (\ref{eq:ly_vs_Re^alph}) and (\ref{eq:ly_vs_Re_dd}). This is a convenient comparison method because the predicted value of $\delta u_f$ in Eq.(\ref{eq:decaying_predict}) is very sensitive to changes in $\lambda(t)$. Using these laws and the statistical properties of the unperturbed field, we can numerically perform the integral in Eq.(\ref{eq:decaying_predict}) without having to measure Lyapunov exponents in the decaying regime. Thus, we predict the behaviour of $E_d(t)$ given its initial value $E_d(10)$. We use a simulation with $N=1024^3$ and we perturb the flow at wavenumber $k=40$. 

\begin{figure}[h!]
    \centering
    \includegraphics[width=8.6cm]{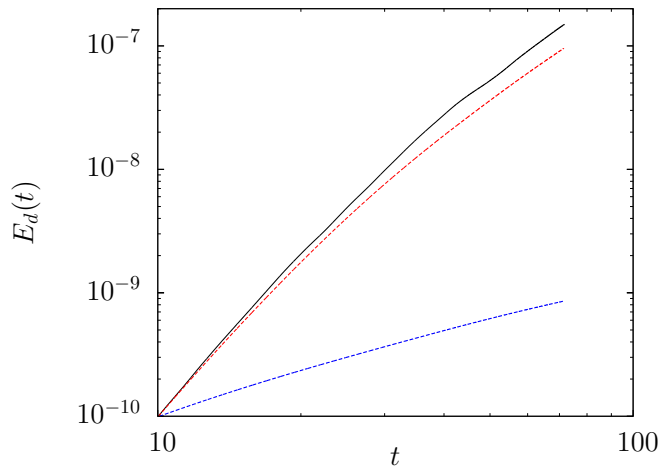}
    \caption{Different predictions of the functional form for $\lambda$. The red dotted line corresponds to the prediction in (\ref{eq:ly_vs_Re_dd}) the blue dotted line to the prediction in (\ref{eq:ly_vs_Re^alph}). The solid black line corresponds to the measured data.}
    \label{fig:decaying}
\end{figure}

Figure \ref{fig:decaying} shows the influence of the different functional forms of $\lambda$ using Eq.(\ref{eq:decaying_predict}) to predict $E_d(t)$ as well as the measured value of $E_d(t)$ for the above mentioned simulation. As can be seen, the prediction given by (\ref{eq:ly_vs_Re_dd}), that uses $T_{E0}$ as the large time scale is the best fit to the measured data. This fit can be improved by using slightly higher values of $D_E$ and $\alpha_E$, which would still be within one standard deviation of the fit for the forced data. This is further evidence that the functional form of $\lambda$ predicted by Eq.(\ref{eq:ly_vs_Re_dd}) is the best fit to data. This has implications for the physical origin of the Eulerian chaos in turbulence, since we cannot derive Eq.(\ref{eq:ly_vs_Re_dd}) from the prediction $\lambda \sim 1/\tau$ consistently with the Kolmogorov theory. As such, chaos in HIT must have some different physical basis. 

\section{Timescales in HIT}
\label{se:timescales_HIT}
Quantities such as energy, Reynolds number, Lyapunov exponents and dissipation rate fluctuate around a mean value during a steady state. We look at the timescales of these time signals produced in our simulations. Especially we look at the fluctuations in the Lyapunov exponent and we compare it with those of energy and dissipation rate. One of the interesting features noted in \cite{mohan2017scaling} is that the fluctuations in the Lyapunov exponents are faster than those of energy or dissipation rate. This can occur due to the perturbation we introduce every steptime $\Delta t$.  Here a more systematic study of this timescale is presented. We observed the autocorrelation of the signal to measure its self-decorrelation time, which we will simply call decorrelation time.
Given the above findings that histograms for some magnitudes reach a stable distribution faster than for others, we may wonder what is the run time necessary to get a stable measurement of different variables depending on the timescale inherent in each time signal. In order to determine such time, it is useful to look at the decorrelation time. 

Previous results show that $E_{ud}$ has a limit to its growth rate which
is roughly equal to $\varepsilon$ \cite{Berera2018}.
We also know that the value of $E_{ud}$ saturates at $2E$ 
As such, the minimum time that it takes for two fields to become completely
decorrelated is $2E/\epsilon = 2T_{E0}$.
This result means that the timescale that is relevant for the energy and other large scale
properties are dependent on $T_{E0}$.

For this purpose, we use the 
autocorrelation in time $\rho_{X,\Delta t}$ for some random variable $X$.
The autocorrelation is related to the covariance $\rho_{X,Y}$, defined as
\begin{align}
\rho_{X,Y} = \frac{E[(X-\mu_X) (Y-\mu_Y)]}{\sigma_X \sigma_Y} \ ,
\end{align}
where $E[...]$ is the expectation, $Y$ is some random variable, $\mu$ is the mean,
and $\sigma$ the standard deviation.
If we then define $Y$ as the value of $X$ after some time $\Delta t$ (in this section $\Delta t$ is not related in anyway to the steptime $\Delta t$ used in previous sections),
then we can define $\rho_{X,\Delta t}$ as equivalent to $\rho_{X,Y}$ here.
In this case, $X$ and $Y$ should have the same statistics, and
so the equation simplifies to
\begin{align}
\rho_{X,\Delta t} = \frac{E[X X_{\Delta t}] - \mu_X^2}{Var[X]} \ ,
\end{align}
where $X_{\Delta t}$ is the value of $X$ after time $\Delta t$, and $Var[...]$ is the variance.

We perform simulations of forced HIT using a value of $\nu = 0.006$ on a lattice of
size $128^3$ with $\varepsilon = 0.1$.
The simulations had $T_{E0} = 5.43$ in simulation time units,
with Re $= 130$.
In this instance, we run it until simulation time 1000, which is over 450 large eddy
turnover times.

\begin{figure}[h!]
    \centering
    \includegraphics[width = 8.6cm]{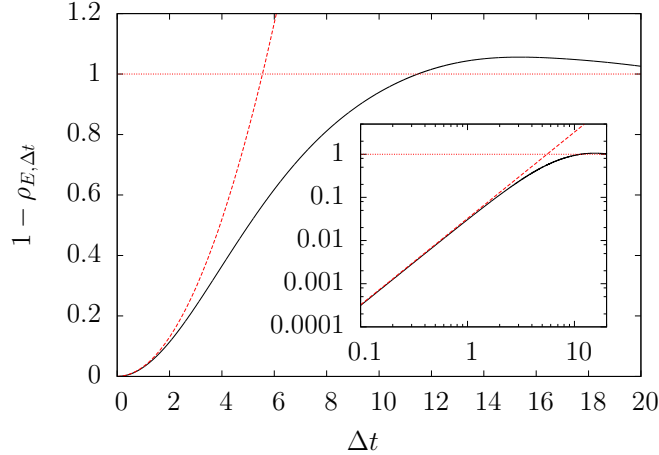}
    \caption{Solid line shows complementary autocorrelation of energy $1- \rho_{E,\Delta t}$
    for Re $= 130$. Dashed line shows scaling with $(\Delta t)^2$, dotted line shows
    constant 1.
    Inset is same as main plot but with log-log axes.}
    \label{fig:EnergyAutoCorr}
\end{figure}

\begin{figure}
    \centering
    \includegraphics[width = 8.6cm]{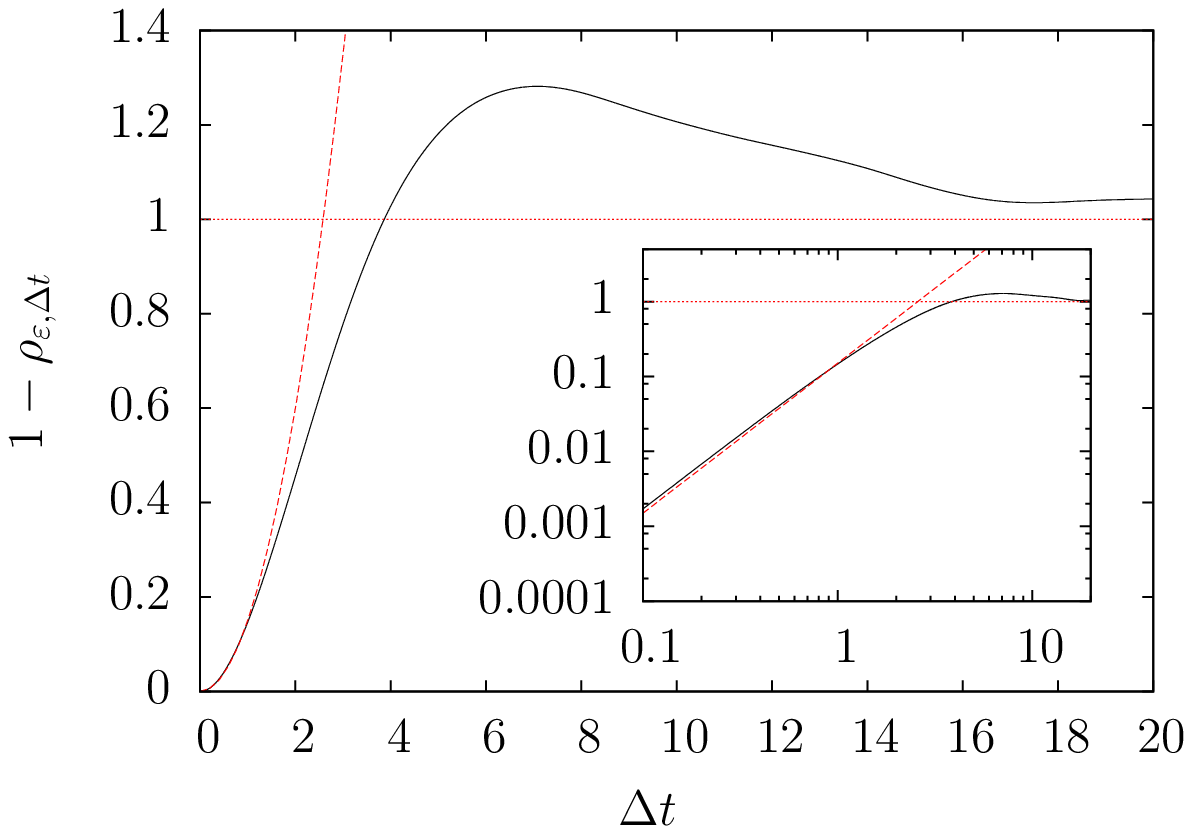}
    \caption{Solid line shows complementary autocorrelation of dissipation
    $1- \rho_{\varepsilon,\Delta t}$
    for Re $= 130$. Dashed line shows scaling with $(\Delta t)^2$, dotted line shows
    constant 1.
    Inset is same as main plot but with log-log axes.}
    \label{fig:DissipationAutoCorr}
\end{figure}

Figure \ref{fig:EnergyAutoCorr} shows $(1- \rho_{E,\Delta t})$ for the simulation
described in the previous paragraph.
For small $\Delta t$ we can approximate $1- \rho_{E,\Delta t} = (\Delta t/T_d)^2$, where we define $T_d$ as the characteristic decorrelation time. In this case, we find that $T_d = 5.55$, which is very close to the measured value of $T_{E0}$.
Similarly, we find that the point at which $\rho_{E,\Delta t} = 0$ is
at $\Delta t = 11.48$, which agrees with our prediction that the time
for two fields to become completely decorrelated is roughly $2T_{E0}$. We expect that the important timescale for measurement may depend on the length scale which is most important for  those  statistics. For  instance, whilst total energy will be dominated by large scale structures, the dissipation should be dominated by the smallest lengthscales. Figure \ref{fig:DissipationAutoCorr} shows $1- \rho_{\varepsilon,\Delta t}$
for the same simulation.
Approximating $1- \rho_{\varepsilon,\Delta t} = (\Delta t/T_d)^2$ gives
$T_d = 2.58$, which is approximately half the value of $T_{E0}$.
Similarly, the time at which $\rho_{\varepsilon,\Delta t} < 0$ is actually less
than $T_{E0}$.

We calculate the autocorrelation for $E(k)$ in order to look
at the dependence on the length scale explicitly.
The plot for $\rho_{E(1),\Delta t}$ looks very similar to those
for $\rho_{E,\Delta t}$.
However, the plots for $k > 1$, even as low as $k = 2$, look more like
the plots for $\rho_{\varepsilon,\Delta t}$, despite the fact that $k = 2$ is
within the forcing range.
All of these plots reach $\rho_{E(k),\Delta t} \leq 0$ quicker than $T_{E0}$
when $k > 1$. For moderately high $k$, there is a $(\Delta t/T_d)^2$ dependence at low $\Delta t$.
But, when we look at the autocorrelation for very high wave number, something
unexpected happens, which is that a new scaling appears.

\begin{figure}[h!]
    \centering
    \includegraphics[width = 8.6cm]{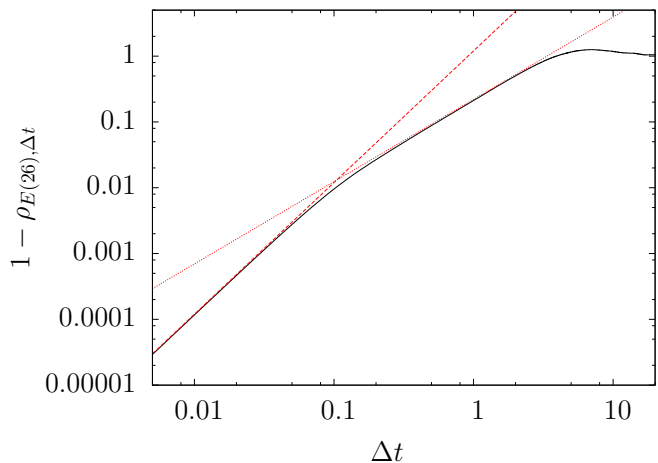}
    \caption{Solid line shows complementary autocorrelation of $E(26)$,
    $1- \rho_{E(26),\Delta t}$
    for Re $= 130$. Dashed line shows scaling with $(\Delta t)^2$, dotted line shows
    scaling with $(\Delta t)^{5/4}$.}
    \label{fig:k26AutoCorr}
\end{figure}

Figure \ref{fig:k26AutoCorr} shows $1- \rho_{E(26),\Delta t}$.
The value $k = 26$ is chosen because it is roughly the inverse of the Kolmogorov
microscale for this simulation.
The new scaling is approximately $(\Delta t)^{5/4}$.
Even with this new scaling, the autocorrelation still becomes decorrelated
before $T_{E0}$.

\begin{figure}[h!]
    \centering
    \includegraphics[width = 8.6cm]{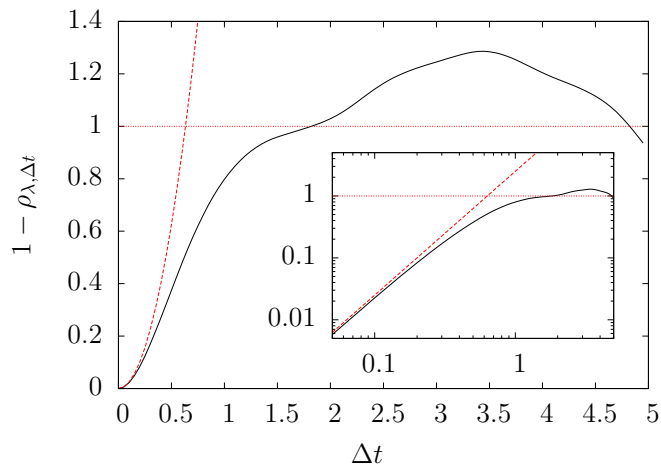}
    \caption{Solid line shows complementary autocorrelation of $\lambda$,
    $1- \rho_{\lambda,\Delta t}$
    for Re $= 130$. Dashed line shows scaling with $(\Delta t)^2$, dotted line shows
    constant 1.
    Inset is same as main plot but with log-log axes.}
    \label{fig:FTLEAutoCorr}
\end{figure}

We now look at the autocorrelation for $\lambda$ itself.
Figure \ref{fig:FTLEAutoCorr} shows $1- \rho_{\lambda,\Delta t}$.
These statistics are measured on a simulation with a much shorter run time
and with FTLE steptime $\Delta t = 0.05$.
We find that $T_d \approx 0.5$ and the time until decorrelation is roughly $T_0/4$.
This is further evidence that the FTLE should reach stable statistics faster
than for instance, the Reynolds number or total energy. This is one of the main results in this paper.
We may expect this longer correlation time for the statistics like total energy, Re, 
and dissipation compared to the Lyapunov exponent just from looking at their
evolution in time. The graphs for Lyapunov exponent fluctuate more wildly.
The correlation between Lyapunov exponents has a much shorter timescale
which is why the statistics of the Lyapunov exponent are more quickly
approximated by a Gaussian. 
However, the Lyapunov exponent becomes decorrelated even faster than the
energy at the Kolmogorov microscale, even though we would expect those
scales to be evolving the fastest in the whole system. This might indicate that the perturbations introduced in the FTLE method have a relevant effect on the Lyapunov exponent self-correlation time. However we cannot consider successive values of $\lambda$ to be uncorrelated since the decorrelation timescale $T_d$ is much larger than the steptime. 

Given the decorrelation time $T_d$, we need to determine the run time $T_r$ necessary to obtain proper statistics for each quantity. It is important to mention that $T_r$ is not the actual run time of the simulation but the time in which the simulation is in the steady state, so initial transients are not considered. 

For this purpose, a useful quantity to look at is the standard error of the mean $\sigma_{\overline{X}}$. We may consider a set of measurements of a quantity $X$, with its own mean $\overline{X}$ and standard deviation $\sigma$. If instead, we consider a subset of $n$ measurements, we can measure a mean value of $X$ ($\overline{X}_{\text{sub}}$) that will usually differ from the real mean value $\overline{X}$. Repeating this procedure many times for different subsets, we can get a distribution of $\overline{X}_{\text{sub}}$. The standard deviation of such a distribution is defined as the standard error of the mean $\sigma_{\overline{X}}$. It is known that for Gaussian distributions, the relation $\sigma_{\overline{X}}/\sigma = 1/\sqrt{n}$ holds, where $\sigma$ is the standard deviation of the entire set, so as $n$ grows, the estimate of the mean improves. 

To associate the value of $n$ to the time signals obtained in our simulations, we divided the entire run time in $n$ segments of length given by the decorrelation time $T_d$, so $n = T_r/T_d$, then, the relation between the standard error of the mean and the standard deviation is now 
\begin{equation}
    \frac{\sigma_{\overline{X}}}{\sigma} = \sqrt{\frac{T_d}{T_r}} \quad ,
    \end{equation}
where $\sigma$ is now the standard deviation of the signal in a simulation with $T_r \rightarrow \infty$. As a rule of thumb, the standard error of the mean should be no greater than $10\%$ of the standard deviation, so a value of $T_r \approx 100 T_d $ will ensure this condition. That is the reason why in the case of measuring Lyapunov exponents, for which $T_d \approx 0.5 \approx T_{E0}/10$, a run time of $T_r = 50 \approx 10 T_{E0}$ is enough to obtain proper statistics, whereas for the energy, that same run time is not enough. Note that the above criteria to determine the run time is uniquely dependent on the decorrelation time, since it is not necessary to measure $\sigma$ and $\sigma_{\overline{X}}$ once we observe that the distribution is approximately Gaussian.

On the other hand, the proper run time needed to obtain proper statistics for the energy is approximately $100 \, T_{E0}$, which is ten times greater than that needed in the case of Lyapunov exponents. The measurement of Lyapunov exponents are the most robust in these simulations. This suggests that the Lyapunov exponents are not only stable to variations in the steptime and lattice size, but also the run time needed to compute them is much faster than for other flow quantities. Thus, using relations such as Ruelle's, that connect the Lyapunov exponent to Reynolds number, is useful to attempt to characterize the flow in a quick and robust way.

\section{MHD}
\label{se:MHD}

\begin{figure}
 \centering
 \subfigure[]{\label{fig:FTLE_histo_MHD}\includegraphics[width=8.6cm]{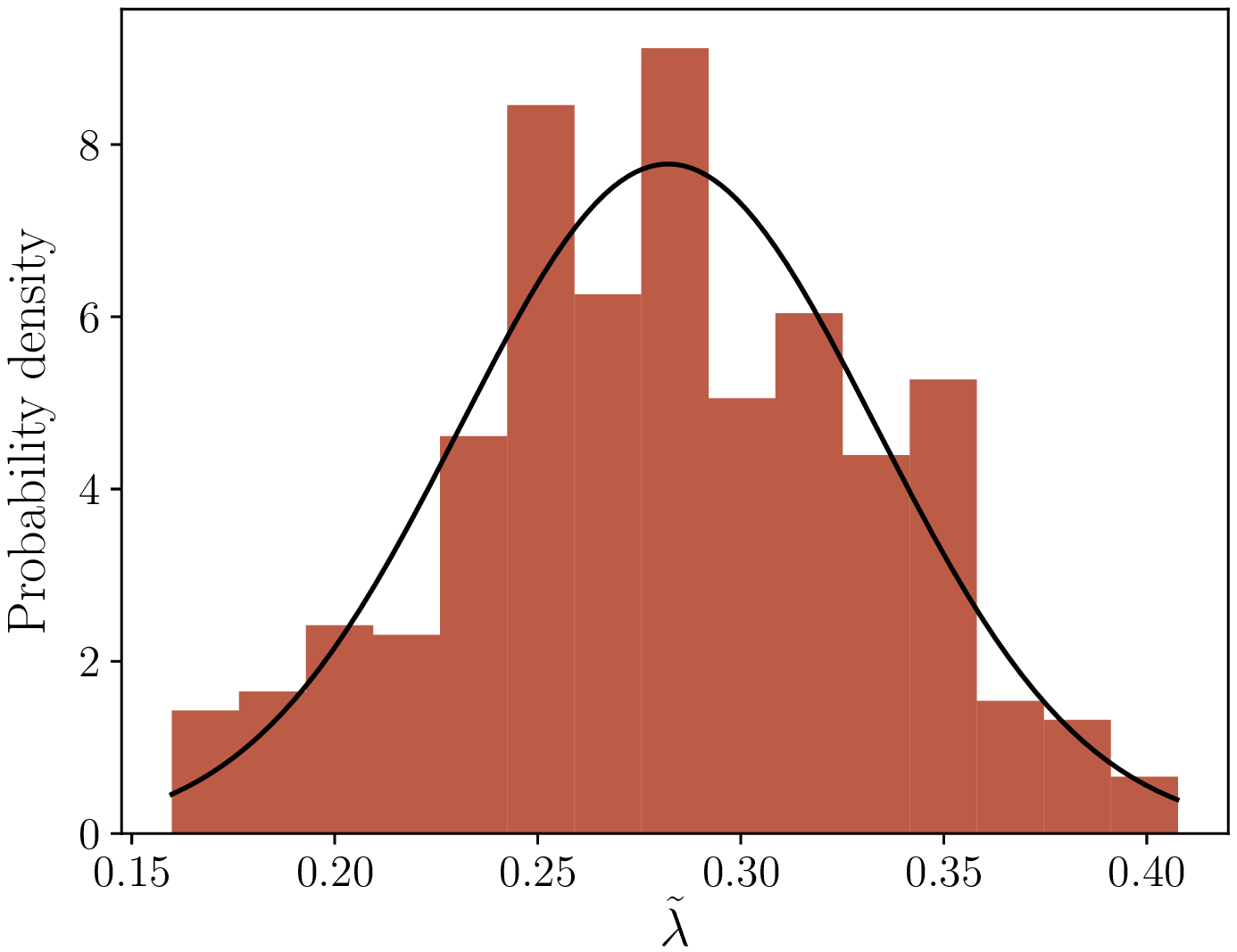}}
 \subfigure[]{\label{fig:histo_Re_MHD}\includegraphics[width=8.6cm]{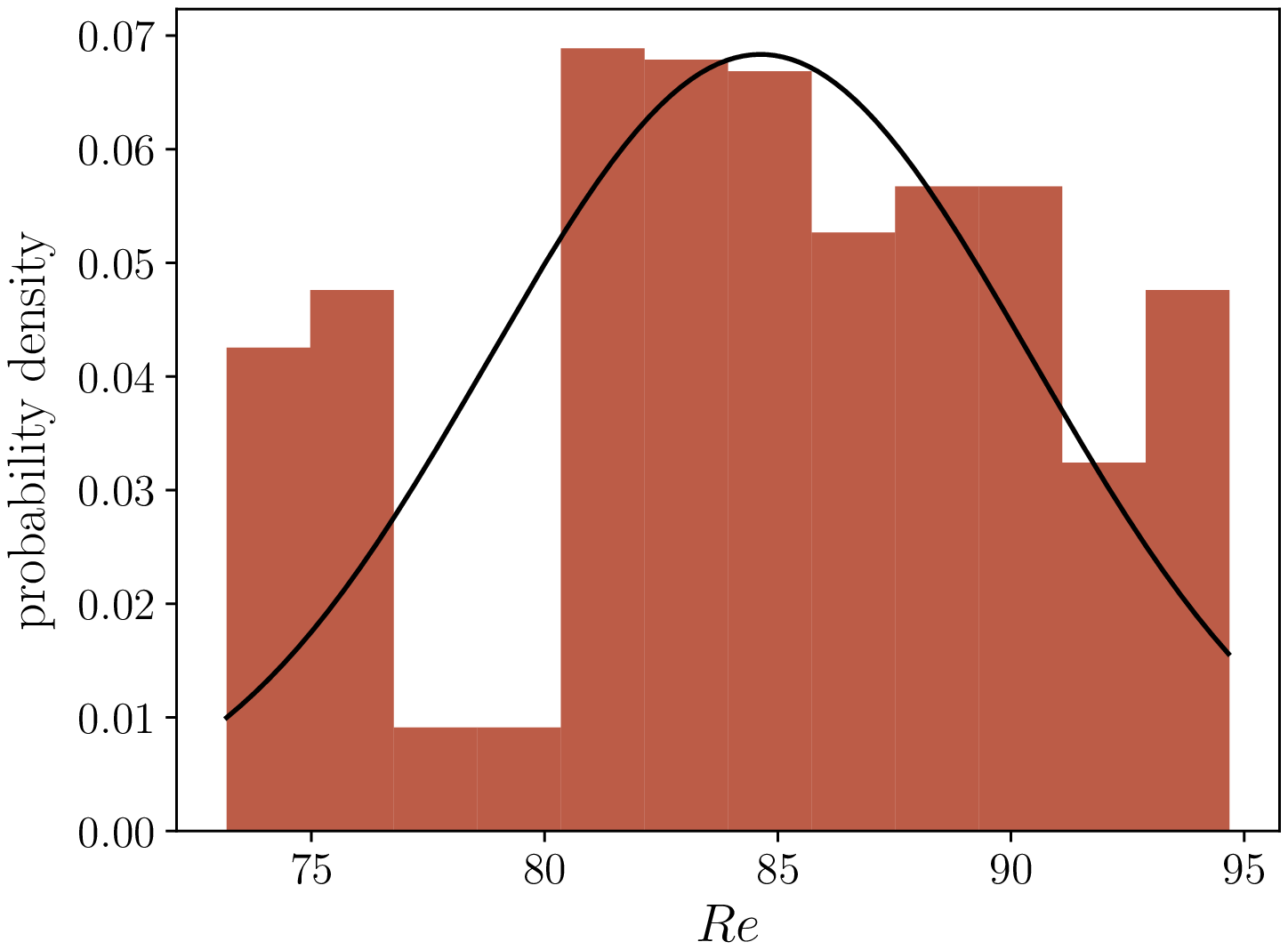}}
\caption{\label{fig:histo_MHD} Distributions of $\lambda$ and Re for the MHD case for one
simulation.}
\end{figure}

\begin{table}[]
    \centering
    \begin{tabular}{ccccccccc}
    \toprule
 $N^3$ &   $\nu$ &  $\varepsilon$ &      Re &  $\lambda$ &  $\sigma_{\lambda}$ &     $T_0$ &  $\tau$ &  $k_{\text{max}}\eta$ \\
\hline
  64$^3$ &  0.0200 &          0.084 &    44.0 &    0.284 &                0.031 &  3.041461 &   0.488 &                  1.98 \\
 128$^3$ &  0.0100 &          0.062 &    84.0 &    0.282 &                0.051 &  2.891003 &   0.402 &                  2.60 \\
 128$^3$ &  0.0080 &          0.055 &   109.0 &    0.275 &                0.046 &  2.875559 &   0.380 &                  2.26 \\
 128$^3$ &  0.0060 &          0.048 &   139.0 &    0.312 &                0.048 &  2.912633 &   0.353 &                  1.89 \\
 128$^3$ &  0.0040 &          0.039 &   201.0 &    0.302 &                0.059 &  2.940972 &   0.320 &                  1.47 \\
 256$^3$ &  0.0020 &          0.032 &   397.0 &    0.336 &                0.068 &  3.022230 &   0.251 &                  1.88 \\
 256$^3$ &  0.0018 &          0.031 &   462.0 &    0.351 &                0.055 &  3.008904 &   0.241 &                  1.75 \\
 256$^3$ &  0.0016 &          0.031 &   500.0 &    0.363 &                0.073 &  2.989567 &   0.227 &                  1.60 \\
 256$^3$ &  0.0014 &          0.031 &   545.0 &    0.382 &                0.114 &  2.938673 &   0.214 &                  1.45 \\
 512$^3$ &  0.0010 &          0.029 &   786.0 &    0.365 &                0.070 &  3.158999 &   0.187 &                  2.31 \\
 512$^3$ &  0.0008 &          0.028 &  1089.0 &    0.414 &                0.049 &  3.205189 &   0.170 &                  1.97 \\
 512$^3$ &  0.0006 &          0.029 &  1342.0 &    0.461 &                0.055 &  2.940171 &   0.145 &                  1.58 \\
 \hline
    \end{tabular}

    \caption{Simulation parameters for 12 MHD simulations. N is the lattice size, $\nu$ is the kinematic viscosity, $\epsilon$ is the kinematic dissipation rate, Re is the Reynolds number, $\lambda$ is the maximal Lyapunov exponent and $\sigma_{\lambda}$ its standard deviation, $T_0$ is the large eddy turnover time, $\tau$ is the Kolmogorov microscale time, $k_{\text{max}}\approx N/3 -1$ is the maximum resolved wavenumber, $\eta = (\nu^3/\varepsilon)^{1/4}$ is the Kolmogorov microscale length and $k_{\text{max}}\eta$ is the resolution.}
    \label{tab:MHD_sim}
\end{table}
Twelve simulations were run using the incompressible MHD equations described in Equations
(\ref{eq:MHDeqns1} - \ref{eq:MHDeqns2}).
These simulations keep magnetic Prandtl number, Pm $ = \nu/\eta$, 
set to 1 for all simulations.
The forcing is set so that the sum of magnetic and kinetic dissipation is
approximately 0.1 and magnetic helicity is kept low.
The FTLE method was used to obtain the Lyapunov exponents. The values of Re ranged between 40 and 1300. Figure \ref{fig:histo_MHD} shows histograms for the distribution of $\lambda$ and
Re obtained from one of the simulations. Table \ref{tab:MHD_sim} shows parameter values for all simulations.
Similar to the hydrodynamic case, whilst the $\lambda$ are well approximated
by a Gaussian distribution, the value for Re is not. As is known, analysis of the chaotic properties of MHD simulations presents noisier results
than the NSE case \cite{ho2019chaotic}.
As well, we find that characteristic quantities and their fluctuations
are less correlated. 
MHD involves more time and length scales than magnetically neutral flows,
given that for each velocity field length and time scale, there is a corresponding time
or length scale for the magnetic field. 
As well, there are new relations between these scales that are possible 
given the presence of magnetic diffusion.
Hence, it is to be expected that some dependencies 
that are obtained from dimensional considerations in hydrodynamics are lost in MHD.  

\begin{figure}[h!]
    \centering
    \includegraphics[width = 8.6cm]{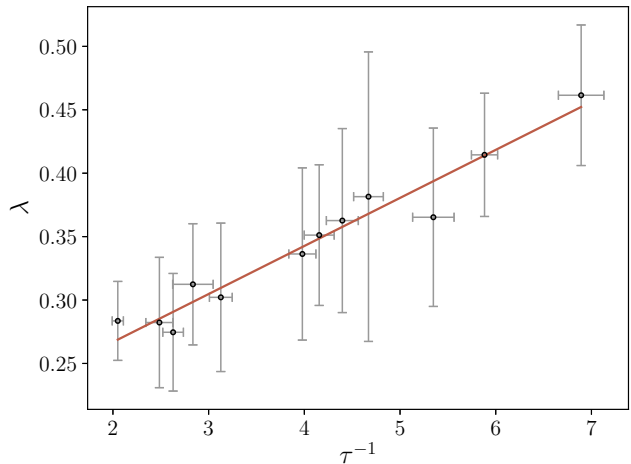}
    \caption{$\lambda$ vs $\tau^{-1}$  (the inverse of Kolmogorov time $\tau = (\nu/\epsilon)^{1/2}$) and linear fit represented by the solid red line. }
    \label{fig:MHD_ly_v_invtau}
\end{figure}

The first relationship to test for MHD simulations is
the Ruelle relation $\lambda \propto 1/\tau$.
Here, $\tau = (\nu/\epsilon)^{1/2}$ is the Kolmogorov time, and the dissipation
is the kinetic dissipation.
Figure \ref{fig:MHD_ly_v_invtau} shows $\lambda$ against $1/\tau$.
A linear fit
$\lambda = A + B \tau^{-1}$ is done. 
The measured values are $A = 0.19 \pm 0.02$ and $B = 0.037 \pm 0.004$. Nevertheless, in \cite{ho2019chaotic} the same plot is presented for a wider range of values of $\tau^{-1}$, showing that for the values of $\tau^{-1}$ we explore, the linear approximation is not accurate whereas for those values that are not explored here the linear approximation becomes reasonable. Exploring a wider range of values to evaluate this relation is not the main goal of
this project and that would require computational cost which is not pursued for this work.  

\begin{figure}
    \centering
    \includegraphics[width = 8.6cm]{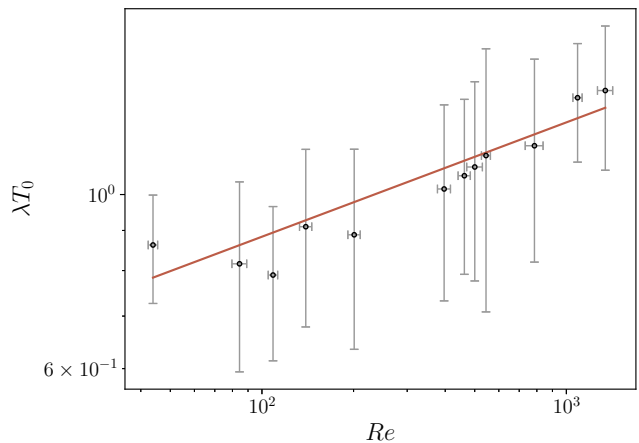}
    \caption{$\lambda T_0$ vs. Re using the FTLE method in a MHD fluid}
    \label{fig:MHD_lyT_v_Re}
\end{figure}

Figure \ref{fig:MHD_lyT_v_Re} shows the relationship between $\lambda$ and Re.
This plot allows us to test the scaling relation in Equation (\ref{eq:ly_vs_Re^alph}).
From the Figure, we measure a value of $\alpha = 0.14 \pm 0.02$,
which is quite far from the value of $\alpha \approx 0.5$ 
found previously for hydrodynamic simulations.
The derivation of Eq.(\ref{eq:ly_vs_Re^alph}) relies strongly on the fact that 
the characteristic quantities of the flow at different scales 
are less ambiguous in hydrodynamics, whilst for MHD, many other quantities such as 
magnetic diffusion or magnetic field length and time scales bring more ambiguity in the dimensional analysis. The correlation observed between $\sigma_{\lambda}$ and Re is very weak, contrary to the hydrodynamic case, where there was a strong correlation as is shown in Figure \ref{fig:DlyT_v_R}.

\begin{figure}
 \centering
\includegraphics[width=8.6cm]{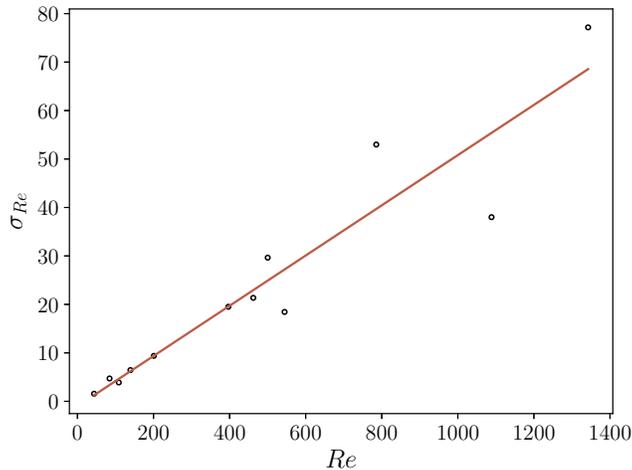}
\label{fig:MHD_DRe_v_Re}
\caption{\label{fig:MHD_error_vs_mean} Distributions of $\lambda$ and Re for the MHD case. No correlation was shown between $\lambda$ and $\sigma_{\lambda}$. The relation between Re and its fluctuations presented a linear relation with a fit given by the solid line.}
\end{figure}

We now compare the dependence of $\sigma_{\lambda}$ on $\lambda$
and $\sigma_{\text{Re}}$ on Re, as was done for hydrodynamics previously.
Figure \ref{fig:MHD_error_vs_mean} shows these relations for our MHD simulations.
There are similarities and differences with the hydrodynamic results.
Unlike for NSE, in MHD the fluctuations in the Lyapunov exponents
are not strongly related to $\lambda$.
However, similar to the NSE case, in MHD the fluctuations of Re are strongly
related to Re itself. From the Figure, a linear fit for the MHD simulations of 
$\sigma_{\text{Re}} = c_3 \text{Re}$ results in a measured value for the slope of $c_3 = 0.052 \pm 0.006$.
This value is of the same order of the analogous slope for hydrodynamics $c_1 = 0.08 \pm 0.01$.

The fact that the relation between Re and its fluctuations are similar 
for both NSE and MHD runs shows that this relation is probably
a characteristic of the forced equations. Although in both MHD and hydrodynamics, the relationship between $\lambda$ and $1/\tau$
is strong, the strong relationship between $\tau$ and Re only holds in hydrodynamics.
Since these relations are not uniquely determined in MHD, 
the behavior of $\lambda$ is not as strongly related to Re. The lack of a uniquely determined relation
may be the reason why the linear relation between $\sigma_{\lambda}$ 
and $\lambda$ is not present in MHD. 

\begin{figure}
    \centering
    \includegraphics[width = 8.6cm]{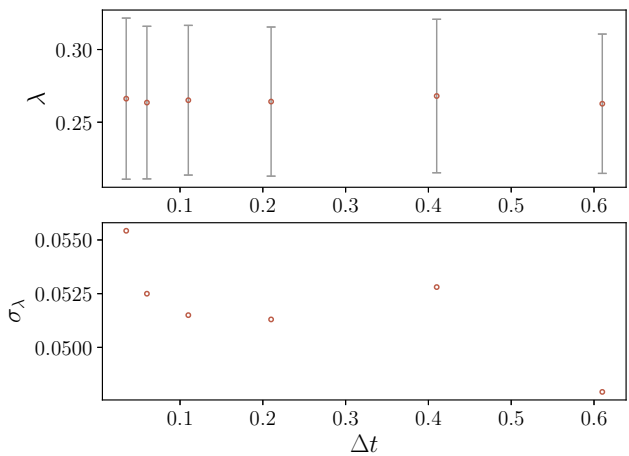}
    \caption{Dependence of $\lambda$ and $\sigma_{\lambda}$ on the steptime $\Delta t$ of the FTLE method for MHD flows.}
    \label{fig:MHD_steptime}
\end{figure}

Similar to hydrodynamics, we test the sensitivity of the MHD FTLE results to changes
in steptime. Figure \ref{fig:MHD_steptime} shows the dependence of $\lambda$ and $\sigma_{\lambda}$ 
on the steptime $\Delta t$ of the FTLE method for one MHD simulation.
As in the hydrodynamic case, both $\lambda$ and $\sigma_{\lambda}$ seem to be stable
even though a very wide range of $\Delta t$ is explored.

\section{Discussion and conclusions}
\label{se:discussions}

We studied Eulerian chaos in homogeneous and isotropic flows evolved using the incompressible Navier-Stokes equations. We observed the finite time Lyapunov exponents and the dependence of fluctuations on other parameters such as Reynolds number, lattice size and steptime in the FTLE procedure. There are four main results which we may take from the analysis done in this paper.

The first is that the FTLEs are approximately Gaussian distributed.
We do not expect that they are fully Gaussian however,
because the statistics on which it depends, and turbulent statistics in general,
are known to be non-Gaussian.

We may try to find out the true distribution of the FTLEs
by relating it to previous results found by analysing HIT with dynamical systems theory. Previously, it has been found that when the NSE are modelled
with a negative damping force, at low Re a relaminarization process occurs 
akin to that found in parallel wall-bounded shear flows \cite{Linkmann2015}.
The lifetime of this process was found to depend super-exponentially on
the Reynolds number.
We can associate the parts of the distribution of the FTLE where $\lambda < 0$
with laminar regions of the state space.
The proportion of the total state space which is laminar can be calculated
using the cumulative distribution function (CDF)
and substitute the dependence on Re of the distribution into this CDF.
If we approximate the FTLE as being totally decorrelated,
we can associate the relaminarization rate with the rate at which the FTLE becomes negative.
The CDF of a Gaussian distribution follows a complementary error function.
Substituting the values in this paper to generate a Re dependence for the relaminarization
rate does not generate the super exponential rates found previously in
turbulence \cite{Linkmann2015}.
Motivated by this finding, we could modify the probability distribution
of the FTLEs and perhaps use some form of generalized extreme value distribution,
which have CDFs which are super exponential in form.
However, the situation may be much more subtle and require further analysis than
the one outlined briefly here.

The second finding relates to the most important time scale for chaos in HIT.
We may begin to question whether the relationship derived by Ruelle is the fundamental
one or whether it is actually a consequence of the relationship in Eq.(\ref{eq:ly_vs_Re_dd}) being the fundamental one.
If Eq.(\ref{eq:ly_vs_Re_dd}) were the fundamental relationship,
it would support the idea that the true timescale for the chaotic properties is $E/\varepsilon$
rather than $L/U$.
The study of the chaotic behaviour in decaying turbulence shows that the evolution of the divergence is quite sensitive to the different functional forms proposed. Numerical data supports the idea that $E/\varepsilon$ is the one that best fit the data. 
We also see that the lattice size is not really a factor, and under-resolved 
simulations also capture the chaotic behaviour. 
As such, the smallest length scales may not be as relevant to the chaos and as shown in previous works \cite{Berera2018,Boffetta2017,mukherjee2016predictability}, these are the scales in which the two field initially close become decorrelated first. Conversely, large scale properties take a longer time to decorrelate and thus are more predictable.  

Third, from the analysis of Figure \ref{fig:Lambda_and_error_vs_steptime}, we see that both the Lyapunov exponent
and its fluctuations are very weakly dependent on the steptime used
in the FTLE method.
Practically, this result means that the time needed to get a measure of this chaotic
property can be achieved at low computational cost.

Fourth, motivated by understanding the chaotic properties
of the system, we analyzed the timescale of the signals of the energy, dissipation rate and Lyapunov exponent in the steady state. We look at the time autocorrelation and we see that the energy only
becomes decorrelated after $2T_{E0}$. However, the small scale properties become decorrelated faster than $T_{E0}$ and the Eulerian maximal Lyapunov exponent becomes decorrelated faster than any of these
other statistics, suggesting that it is the most robust to measurement in simulations, reaching a faster stability. 
From this analysis we obtain a useful rule that determines the run time $T_r$ needed to obtain proper statistics of a quantity by considering the decorrelation time in its signal $T_d$, being $T_r\approx 100 \, T_d$. For instance, we see that for the total energy, averages need to be
taken for run times $T_r \approx 100 T_{E0}$, whereas for other properties such as Lyapunov exponents or dissipation rate, shorter run times can be used. This run time rule we give in relation to the self correlation time is useful not only when looking at chaotic properties of the flow but also for any other study that needs to measure quantities in the steady state.

The robustness of the results when measuring the maximal Lyapunov exponent suggest
that it is a stable statistical property of the system. 
We also see that it reaches stability faster than other statistical properties
of the system and our analysis cements it as an extra parameter of HIT which
is important in understanding the interesting inter-scale dynamics inherent in turbulent
flows. This could have further applications in studies that use the Lyapunov exponent to characterize the turbulent flow as in \cite{nastac2017lyapunov}. Furthermore, if we learn the relation between the Lyapunov exponent and other flow quantities, we could use such an exponent to determine these other quantities in a faster and more stable way. For instance, Ruelle's relation relates $\lambda$ and Re and this can be used to estimate Re in a shorter run time. 

Finally, in this paper some of the FTLEs' properties measured for hydrodynamic flows were measured for the case of isotropic and incompressible MHD flows. The distribution of FTLEs is approximately Gaussian as in the case of NSE. The remarkable stability of FTLEs against variations in the steptime is also present in MHD flows. Nevertheless, the dependence on Reynolds number is not the same as in the case of NSE. Ruelle's relation is derived from assuming that the Kolmogorov time is the smallest timescale in the system. Whereas this is true for incompressible NSE, other characteristic timescales are present in the smallest scales in MHD that may break this relation.

\section*{Acknowledgements}
This work used the Cirrus UK National Tier-2 HPC Service at EPCC
funded by the University of Edinburgh and EPSRC (EP/P020267/1).
R.D.J.G.H. was supported by the U.K. Engineering and Physical
Sciences Research Council (EP/M506515/1) and A.A. was supported by
the University of Edinburgh. A.B. acknowledges funding from the U.K.
Science and Technology Facilities Council.

\bibliographystyle{apsrev4-1}
\bibliography{sample.bib}{}

\end{document}